\documentclass[a4paper,fleqn,usenatbib]{mnras}
\usepackage{txfonts}
\usepackage[T1]{fontenc}
\usepackage{ae,aecompl}
\usepackage{graphicx}	
\usepackage{amssymb}	

\usepackage{wasysym}

\usepackage{multicol}

\usepackage{rotating}
\usepackage{hyperref}
\usepackage{url}
\usepackage[all]{hypcap}
\hypersetup{
    bookmarks=true,         
    unicode=false,          
    pdftoolbar=true,        
    pdfmenubar=true,        
    pdffitwindow=false,     
    pdfstartview={FitH},    
    pdftitle={My title},    
    pdfauthor={Author},     
    pdfsubject={Subject},   
    pdfcreator={Creator},   
    pdfproducer={Producer}, 
    pdfkeywords={keyword1} {key2} {key3}, 
    pdfnewwindow=true,      
    colorlinks=true,        
    linkcolor=blue,         
    citecolor=blue,         
    filecolor=magenta,      
    urlcolor=cyan           
}

\title[Triggering of circular--cum--parallel ribbon flare]{Evolutionary stages and triggering process of a complex eruptive flare with circular and parallel ribbons}

\author[N.C. Joshi et al.]{
Navin Chandra Joshi,$^{1,2}$\thanks{E-mail: ncjoshi@srmuniversity.ac.in, njoshi98@gmail.com}
Bhuwan Joshi,$^{1}$
and Prabir K. Mitra$^{1,3}$
\\
$^{1}$Udaipur Solar Observatory, Physical Research Laboratory, Udaipur 313 001, India\\
$^{2}$Department of Physics, SRM University, Delhi-NCR, Sonepat-131029, Haryana, India; ncjoshi@srmuniversity.ac.in\\
$^{3}$Department of Physics, Gujarat University, Ahmedabad 380 009, India}
\date{Accepted 2020 November 2. Received 2020 October 23; in original form 2020 March 21}

\pubyear{2020}

\begin{document}
\label{firstpage}
\pagerange{\pageref{firstpage}--\pageref{lastpage}}
\maketitle

\begin{abstract}

We report multi--wavelength study of a complex M--class solar eruptive flare that consists of three different sets of flare ribbons, viz. circular, parallel, and remote ribbons. Magnetic field modeling of source active region NOAA 12242 exhibits the presence of 3D null--point magnetic topology which encompasses an inner bipolar region. The event initiates with the faint signatures of the circular ribbon along with remote brightening right from the pre--flare phase which points toward the ongoing slow yet persistent null--point reconnection. We first detected flux cancellation and an associated brightening, which are likely signatures of tether-cutting reconnection that builds the flux rope near the polarity inversion line (PIL) of the inner bipolar region. In the next stage, with the onset of M8.7 flare, there is a substantial enhancement in the brightening of circular ribbon which essentially suggests an increase in the rate of ongoing null--point reconnection. Finally, the eruption of underlying flux rope triggers ``standard flare reconnection" beneath it producing an abrupt rise in the intensity of the parallel ribbons as well as enhancing the rate of null--point reconnection by external forcing. We show that within the the fan dome, the  region with magnetic decay index n>1.5 borders the null-point QSL. Our analysis suggests that both the torus instability and the breakout model have played role toward the triggering mechanism for the eruptive flare. This event is a nice example of the dynamical evolution of a flux rope initially confined in a null-point topology, that subsequently activates and erupts with the progression of the circular--cum-- parallel ribbon flare. 
\end{abstract}

\begin{keywords}
Sun:flare -- Sun:activity -- Sun:X--rays, gamma rays -- Sun:magnetic fields -- Sun:coronal mass ejections (CMEs)
\end{keywords}



\section{Introduction}
\label{}

Solar flares occur due to transient and explosive release of magnetic energy in the solar corona. This magnetic energy is stored in the highly sheared and twisted magnetic field associated with active regions (ARs). The released energy is converted to heat the plasma as well as to accelerate the electrons and other particles \citep[]{Benz08,Shibata11}. 

Various theories/models have been proposed to explain solar flares and associated eruptions \citep{Forbes00,Lin03,Schmieder13}. In the well known two--dimensional (2D) ``CSHKP" model of eruptive solar flares \citep{Carmichael64,Sturrock66,Hirayama74,Kopp76}, the reconnection occurs underneath the erupting filament or the flux rope in a vertical current sheet. The eruption of the filament or the flux rope proceeds to give a coronal mass ejection (CME). The flare arcades form underneath the vertical current sheet. This model also successfully explains the formation of parallel flare ribbons at chromospheric heights along either side of polarity inversion line (PIL). 

The standard flare model does not explain how the eruption is triggered in the first place. Further, the explanations for various three--dimensional (3D) aspects of the eruptive flares are beyond the scope of the standard flare model, such as shear in the flare loops \citep[e.g.,][]{Aulanier12}, converging motions of flare ribbons and X-ray footpoint sources \citep[e.g.,][]{bJoshi09,Joshi2017} and, J--shaped ribbons \citep[e.g.,][]{Chandra09,Joshi2017}. To incorporate these complex observational features \citet{Aulanier12,Janvier13} further described a 3D model based on numerical simulation. This is a magnetohydrodynamics (MHD) model which considers a flux rope that lies in a sigmoidal AR \citep[][]{Savcheva12}. During the build--up phase of the flux rope, the current sheet forms underneath it. The flux rope undergoes eruption due to the torus instability \citep[e.g.,][]{kliem06,Aulanier10}. The magnetic reconnection occurs in the current sheet beneath the erupting flux rope. The current layers are associated with the quasi--separatrix layers \citep[QSLs;][]{Titov99}, which are the regions where the magnetic connectivity has strong gradients. The nature of the reconnection within the QSLs is of a slipping type where the magnetic field lines continuously exchange their connectivity with the neighboring field lines. As a result, the apparent slipping motions of the field lines are observed \citep[e.g.,][]{Dudik16}.

Circular ribbon flares occur in a 3D magnetic configuration and consist of ribbons that present circular or semi--circular morphology \citep[e.g.,][]{Masson09,Reid12,Wang12,Jiang13,Sun13,Vemareddy14b,Joshi15,Joshi17,Hernandez-Perez17,Xu17,Li18,Zhong19}. 
Circular ribbon flares exist because of the reconnection in a 3D null--point topology \citep[e.g.,][]{Torok09,Masson09,Wang12,Sun13,Joshi17}. The photospheric magnetogram observations for such a null--point topology show a central compact polarity surrounded by a polarity of opposite sign \citep[e.g.,][]{Sun13,Joshi17,JoshiN18}. In corona, it involves a set of closed field lines (i.e., the lines enclosed below the fan) connecting the central compact polarity to the surrounding polarity. In this configuration, we further recognize field lines over the fan, which join the surrounding polarity to the far away remote region of polarity with the same sign as that of central compact polarity. In this topology, the fan itself is a separatrix surface and the spine is a singular line that intersects the fan at the null--point. Reconnection can occur between the magnetic fluxes below and above the fan surface. When magnetic fields reconnect at null--point, they pass through the fan separatrix surface and jump from the inner to the outer connectivity domain. Particles accelerated at the reconnection site (i.e., the null--point) propagate along the fan field lines which are anchored close to the fan surface, hence the formation of the circular ribbon. Numerical simulations of this type of 3D reconnection and circular ribbon flares have also been explored in some of the studies \citep[e.g.,][]{Masson09,Pariat09,Torok09,Jiang13,Nayak19}. 
Here, we also highlight some interesting examples of the null--point reconnection that involves external forcing below the null point e.g., filament and/or the flux rope eruption from inside the fan--dome \citep{Jiang13,Jiang14,Sun13,Liu15,Joshi15,Joshi17,Joshi18}; shearing of the dome field lines \citep[e.g.,][]{Vemareddy14b,Xu17,Li17}.

The solar flares which are associated with the eruption of the filaments and/or flux ropes are known as eruptive solar flares. Two representative models -- tether--cutting and breakout -- invoke magnetic reconnection as the triggering mechanism of the eruption of the flux rope. The tether--cutting scenario has been suggested long back \citep{Sturrock84,Moore92} and extended later on by \citet{Moore01}. According to this model, the initial low lying tether--cutting magnetic reconnection at the PIL is responsible for the formation of the flux rope as well as its subsequent destabilization. Several observational studies provide substantial support for the tether cutting model to explain the flux rope eruptions \citep[e.g.,][]{Yurchyshyn06,Chifor07,Raftery10,Fan12,Liu13,Joshi17,Joshi18,Mitra2020a}. The investigations of temporal and spatial evolution of the X-ray emission during the precursor phase together with the activation of underlying flux rope structure (e.g., filament and hot channel) are considered as important signatures for the tether-cutting magnetic reconnection \citep[e.g.,][]{Chifor07,Kim08,Joshi11,JoshiB2013,Mitra2019,Sahu20}. The magnetic reconnection associated with the growth and evolution of flux rope along with associated flare loops during an eruptive flare has been substantiated in the standard flare model in 3D \citep[][]{Aulanier12,Aulanier13,Janvier13}. Here the current sheet forms inside the sheared arcade and magnetic reconnection builds--up the flux rope. As the flux rope forms and rises, the current sheet extends due to the elongation of the flux rope. Thus the role of magnetic reconnection forming the flux rope in 3D model can be considered similar to the tether--cutting process.

Breakout model of the solar eruption was proposed by \citet[][]{Antiochos99} and it relies on resistive instability. They began with quadrupolar magnetic configuration having a single null point in 2.5 dimensional (2.5D) axisymmetric spherical geometry. The model recognizes that the early reconnection (i.e., breakout process) at the magnetic null, lying well above the core field region, can trigger the eruption of core fields. The breakout model was further manifested in 3D null point configuration \citep[][]{Lynch08,Lynch09,Wyper17}.
The 3D configuration typically consists of a fan separatrix surface between two different flux system, spine line and a 3D magnetic null point \citep[e.g.,][]{Lynch08,Lynch09}. The reconnection occurs at 3D magnetic null point \citep[e.g.,][]{Aulanier00,Sun13,Liu15,Xu17,Li18,Wyper17}. Observational evidences for the breakout model has been identified in various studies \citep[e.g.,][]{Sterling01,Manoharan03,Gary04,Sterling04a,Dang05,Pohjolainen05,Joshi07,Aurass11,Chen16,Mitra18}.
\cite{Aulanier00} observed a typical fan--spine type magnetic configuration with a 3D null--point and interpreted the reconnection at the magnetic null--point to be a ``magnetic breakout" process.  \citet{Sun13} and \citet[][]{Xu17} conducted multiwavelength studies of circular ribbon flares and compared the flare morphology with that obtained from magnetic field extrapolation. Those studies identified null-point reconnection, suggesting that the triggering mechanism of the eruption may be the breakout. 

The contemporary studies recognize the eruption of twisted flux ropes by ideal instabilities, viz., kink \citep[e.g.,][]{Torok04} and torus \citep[e.g.,][]{Torok07} instabilities. In the context of present study, we briefly discuss the latter one. Torus instability occurs when the flux rope reaches a region where the decay index is greater than 1.5 i.e., the critical value of decay index \citep[e.g.,][]{Torok07,Aulanier10}. However, some studies point toward a range in the decay index rather than a single critical value, such as, 1.5--1.75 \citep[e.g.,][]{Liu08,Kliem13} and 1.3--1.5 \citep[e.g.,][]{Zuccarello15}.

Recently, few cases of typical flares that consist of two sets of ribbons -- two parallel and one circular ribbon -- have also been reported {\citep{Joshi15,Carley16,Joshi17,Devi20}}. The first observation of this kind was reported by \citet{Joshi15} where they proposed a scenario in which two stages of magnetic reconnection explained the formation of two different sets of flare ribbons.
Later on, \citet{Carley16} studied the same event using Nan\c{c}ay radioheliograph radio data and supported the scenario. These types of events occurred in a typical 3D null--point magnetic configuration with multiple steps in the activation and eruption of the flux rope \citep{Joshi15,Joshi17}. In all these cases, we find that the flux rope eruption and associated standard reconnection occur prior to the null--point reconnection. Regardless of the timing in the evolution of circular and parallel flare ribbons, the topology is almost always the same: a flux rope is enclosed below the fan surface of a null--point topology. In the present work, we study a complex circular ribbon flare during which parallel ribbons are also formed within the extension of the circular ribbon after an interval of about 30 min. Further, the circular ribbon flare was eruptive in nature with clear signatures of flux rope activation and eruption in the source region which eventually resulted into a CME. The event under investigation occurs in the AR NOAA 12242 on 2014 December 17. The event further gives us an opportunity to explore the triggering mechanism of the eruption in the light of contemporary theoretical understandings. The overall organization of the paper is as follows. Section~\ref{sec2} shows the description of the observational data set and the related instruments used for this work. The magnetic field configuration over the AR and the changes of the magnetic field near the inner PIL are discussed in Section~\ref{sec3}. Section~\ref{sec4} describes the multi--wavelength analysis of this event including the temporal evolution of flare (Section~\ref{sec4.1}), the characterization of various phases of the event and the discussion of associated CME (Section~\ref{sec4.2}), and the analysis of RHESSI X--ray data (Section~\ref{sec4.3}). The main results and discussions are presented in Section~\ref{sec5} and conclusions in Section~\ref{sec6}.


\section{Observational Data Set}
\label{sec2}

Multi--wavelength data set is used to analyze this flare event. We mainly use data from Atmospheric Imaging Assembly \citep[AIA;][]{Lem12}, Helioseismic and Magnetic Imager \citep[HMI;][]{Schou12}, \textit{the Geostationary Operational Environmental Satellite (GOES)}, and {\it the Reuven Ramaty High Energy Solar Spectroscopic Imager} \citep[\textit{RHESSI};][]{Lin02} instruments. AIA observes the Sun in seven Extreme Ultraviolet (EUV) (i.e., 304, 131, 171, 193, 211, 335, and 94 \AA), two Ultraviolet (UV) (1600 and 1700 \AA), and one white light wavelengths. It has a temporal cadence of 12 s for EUV images, 24 s for UV images and 1 hr for white light images with the pixel size of $\rm 0.6\arcsec$. AIA 131 \AA\ data is used to study the pre--eruptive phase, while the AIA 94 \AA\ data is used to study the eruption dynamics of the event. AIA 1600 and 304 \AA\ wavelength images are utilized to investigate the flare ribbons and remote brightening dynamics (RB). The line--of--sight photospheric magnetogram having the temporal cadence of 45 s with the pixel size of $\rm 0.5\arcsec$ is observed by HMI. Both AIA and HMI are the instruments on--board \textit{the Solar Dynamics Observatory} \citep[\textit{SDO};][]{Pesnell12}. Integrated full disk X--ray flux at 1.0--8.0 \AA\ and 0.5--4.0 \AA\ wavelength bands are observed by {\it GOES}. X--ray observations at energy bands ranging from 3 keV to 50 keV are obtained from {\it RHESSI}. The image reconstruction is done with the CLEAN algorithm using front detector segments only. Out of 9 detector segments, segments 3--8 (excluding 7) are taken into account. To study the morphology of the associated CME we use the Large Angle and Spectrometric Coronagraph \citep[LASCO;][] {Brueckner95} on board the {\it Solar and Heliospheric Observatory} \citep[\textit{SOHO};][]{Domingo95}.

To investigate the coronal magnetic field topology over the AR, we use potential--field source--surface \citep[PFSS,][]{Wang1992,Schrijver03} as well as the optimization based non--linear force free field \citep[NLFFF,][]{wiegelmann06,wiegelmann12} extrapolation methods. PFSS is an IDL--based algorithm which is available in the SolarSoftWare package (SSW\footnote{\url{http://www.lmsal.com/solarsoft/}}). As boundary condition for the NLFFF extrapolation, we used photospheric vector magnetograms from the $\textit{hmi.sharp\_cea\_720s}$ series of \textit{SDO}/HMI which were preprocessed as described in \citet{wiegelmann06}. 
The input parameters for preprocessing i.e., $\nu$, $w_{los}$, $w_{trans}$, $\mu_1$, $\mu_2$, $\mu_3$, $\mu_4$ \citep[see Section 4.2.1 in][]{Mitra2020b}, were chosen to be 0.01, 1, $\frac{B_{trans}}{max(B_{trans})}$, 1, 1, 0.001 and 0.01, respectively.
The extrapolations were done in a Cartesian volume of $510\times 180\times 180$ pixels with resolution 2$\arcsec$ which translates to a physical volume of $\sim$740$\times 260\times 260$ Mm$^{3}$. 
For visualization of the NLFFF lines, we use Visualization and Analysis Platform for Ocean, Atmosphere, and Solar Researchers software \citep[VAPOR\footnote{\url{https://www.vapor.ucar.edu}},][]{Clyne2007}. 
Using the extrapolated magnetic field, we calculated the net Lorentz force ($|\vec{J}\times\vec{B}|$), current-weighted angle between the modelled magnetic field and electric current density, $\theta_J$, and the volume-averaged fractional flux, $<|f_i|>$ \citep{DeRosa15} which can be considered to
assess the quality of the NLFFF extrapolation. The corresponding values\footnote{For a perfectly solenoidal and force free magnetic field, the values of the parameters assessing the quality of the extrapolation, such as, $|\vec{J}\times\vec{B}|$, $\theta_J$, $<|f_i|>$ should be 0. However, since the code essentially works on numerical calculations using real magnetogram observations, non-zero finite values of these parameters are expected. In general, NLFFF solutions returning the values $|\vec{J}\times\vec{B}|\lesssim 10\times 10^{-3}$, $\theta_J\lesssim10^{\circ}$, $<|f_i|>~\lesssim 20\times 10^{-4}$, are considered as good NLFFF solutions \citep[see, e.g.,][]{DeRosa15,Thalmann19}.} were found to be $5.54\times 10^{-3}$, $8.98^{\circ}$ and $3.92\times 10^{-4}$. The Quas--Separatrix Layer (QSL) is calculated within the extrapolation volume by employing the code developed by \citet{Liu2016}.


\section{Evolution and configuration of magnetic field in AR 12242}
\label{sec3}

\begin{figure*}
\vspace*{1cm}
\centerline{
	\hspace*{0\textwidth}
	\includegraphics[width=1.1\textwidth,clip=]{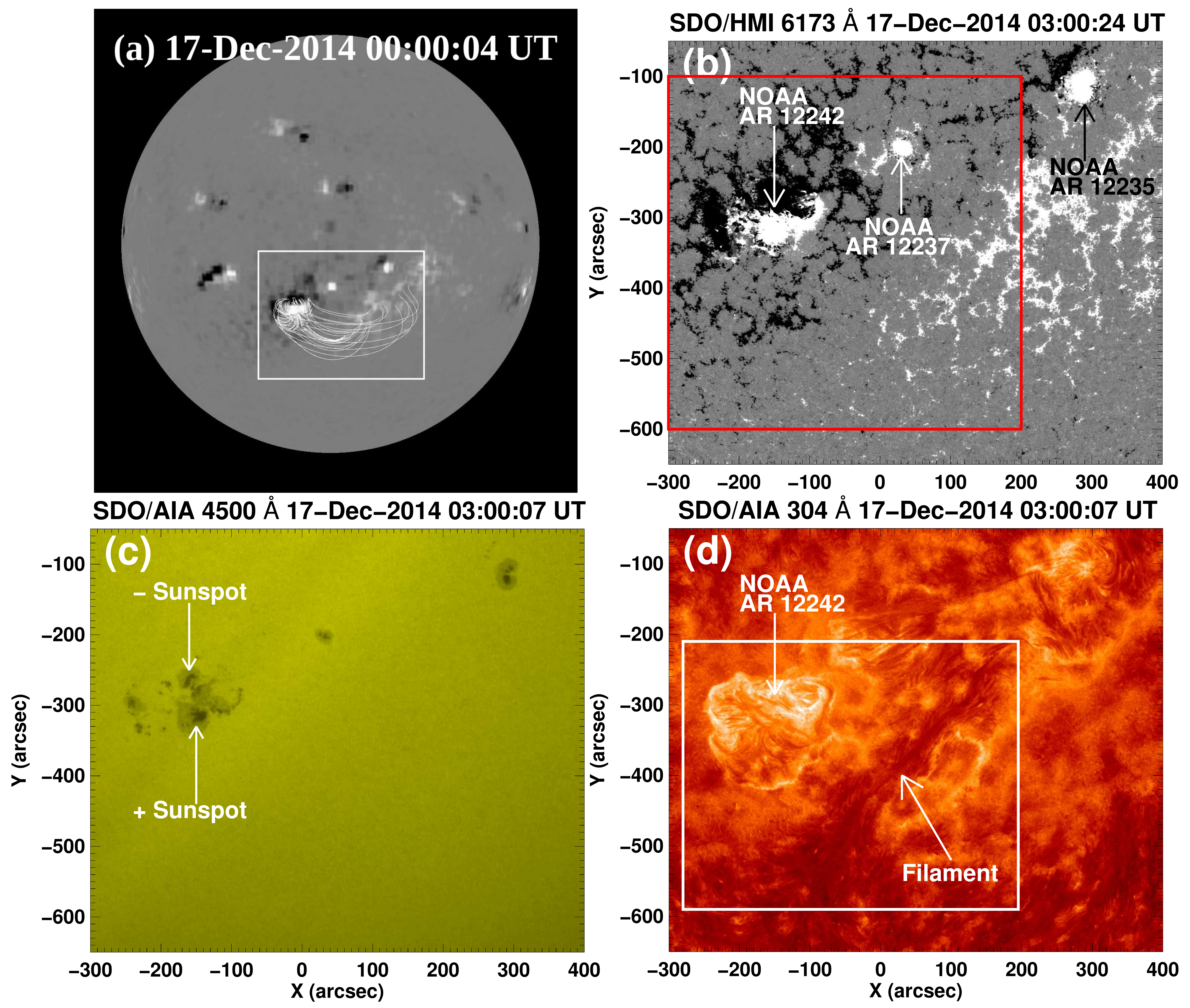}
	}
\vspace*{0cm}
\caption{(a) Full disk \textit{SDO}/HMI line--of--sight photospheric magnetogram at 00:00:04 UT on 2014 December 17 with magnetic field lines obtained from PFSS extrapolation method. (b) \textit{SDO}/HMI line--of--sight photospheric magnetogram on 2014 December 17 at 03:00:24 UT. The red box over the panel (b) represents the field--of--view of Figure~\ref{fig2}(a). (c) \textit{SDO}/AIA 4500 \AA\ image at 03:00:07 UT on 2014 December 17. (d) \textit{SDO}/AIA 304 \AA\ image on 2014 December 17 at 03:00:07 UT, showing the region of flaring activity. White box shows the area used to calculate the \textit{SDO}/AIA intensity profiles shown in Figure~\ref{fig4}(b).}
\label{fig1}
\end{figure*}

On 2014 December 17, AR NOAA 12242 was located at the south--west quadrant of the Sun with heliographic coordinates S19W02. Figure~\ref{fig1}(a), shows the full disk \textit{SDO}/HMI magnetogram at 00:00:04 UT on 2014 December 17 with PFSS extrapolated field lines over the AR 12242. The zoomed image of the line--of--sight (LOS) photospheric magnetogram corresponding to the box in Figure~\ref{fig1}(a) is shown in Figure~\ref{fig1}(b). The magnetic configuration of the AR is very complex and classified as $\rm \beta\gamma\delta$. The AR 12242 mainly consists of two large (indicated by arrows in Figure~\ref{fig1}(c)) and several small sunspots. The northern and southern sunspots in AR 12242 correspond to the negative and positive polarities, respectively (cf. Figures~\ref{fig1}(b) and~\ref{fig1}(c)). The AR structure in the \textit{SDO}/AIA 304 \AA\ wavelength at 03:30:07 UT on 2014 December 17 can be seen in Figure~\ref{fig1}(d). We note the existence of a large filament on the west side of the AR (shown by the arrow in Figure~\ref{fig1}(d)).

\begin{figure*}
\vspace*{1cm}
\centerline{
	\hspace*{0\textwidth}
	\includegraphics[width=1.0\textwidth,clip=]{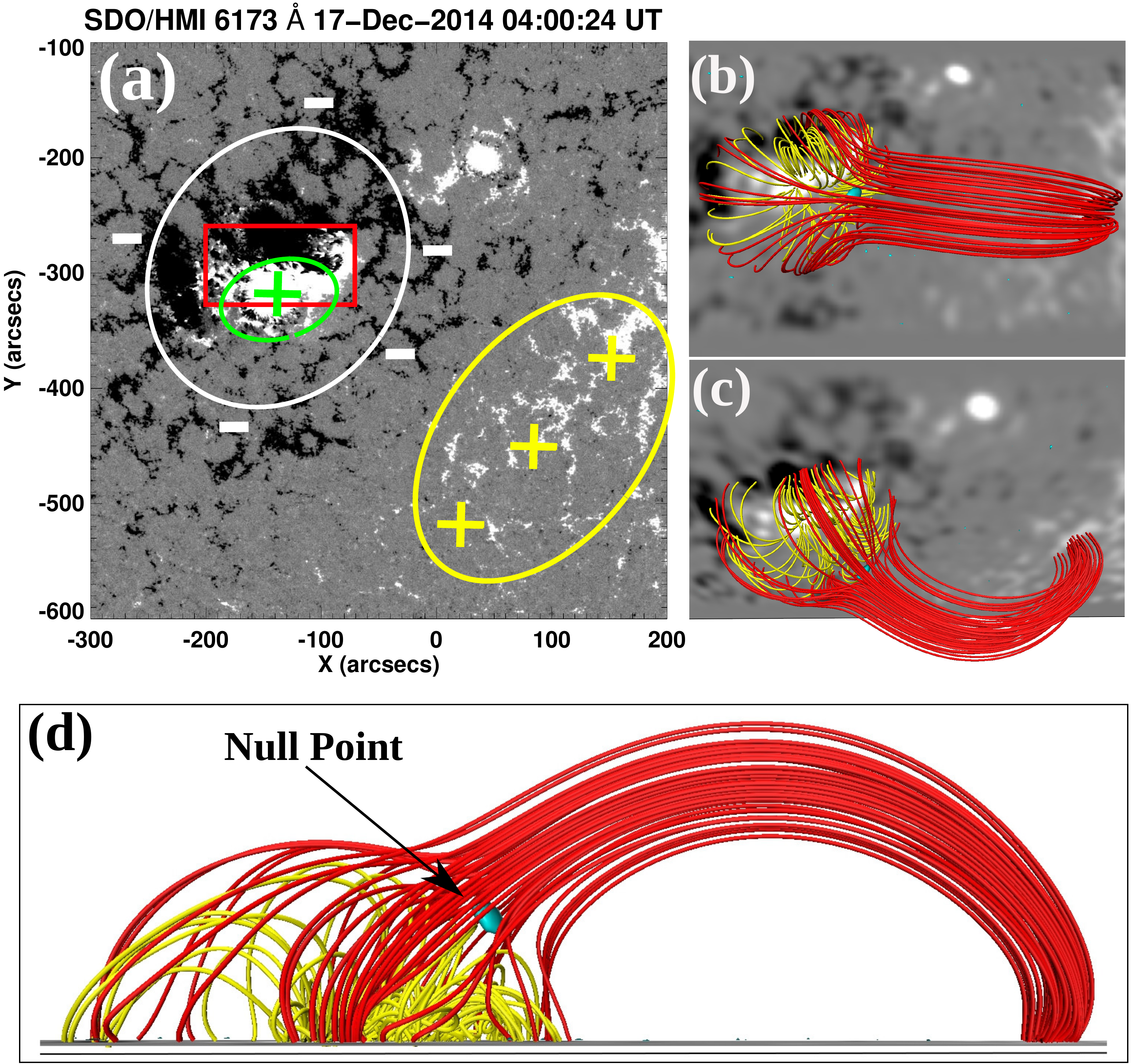}
	}
\vspace*{-0cm}
\caption{(a) \textit{SDO}/HMI line--of--sight photospheric magnetogram at 04:00:24 UT on 2014 December 17. The central positive polarity and the surrounding negative polarity regions are represented by the green and white ellipse, respectively. The flux rope formation and remote brightening regions are marked by the red box and yellow ellipse, respectively. ((b)--(d)) Zoomed view of the coronal field lines showing the null--point topology in three different projections. The yellow and red lines represent the inner dome and outer dome lines, respectively. The cyan patch lying in between the yellow and red field lines in panels (b)--(d) represents an isosurface of 4 Gauss that consists a magnetic null--point.}
\label{fig2}
\end{figure*}

\begin{figure*}
\vspace*{1cm}
\centerline{
	\hspace*{0\textwidth}
	\includegraphics[width=1.0\textwidth,clip=]{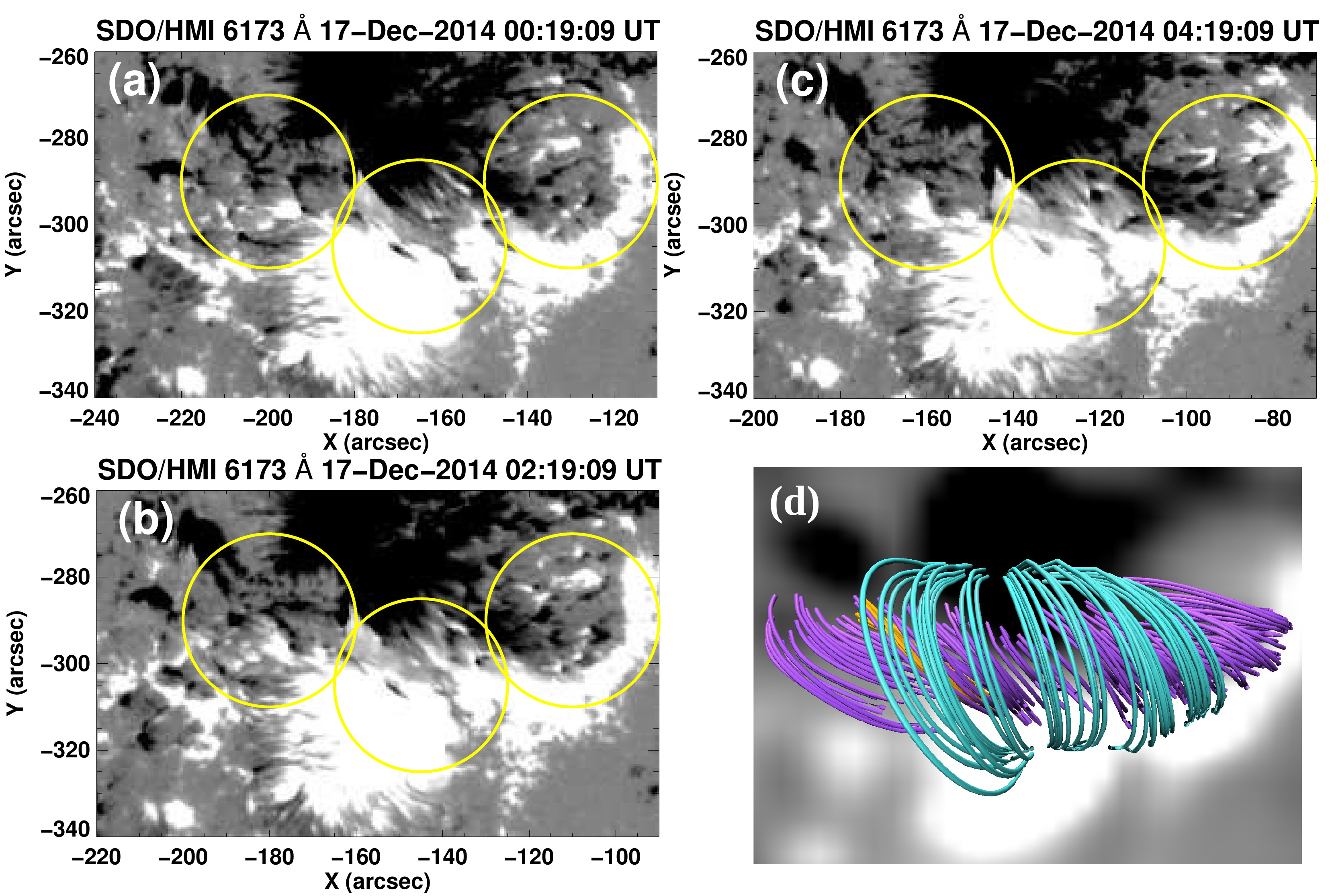}
	}
\vspace*{0cm}
\caption{((a)--(c)) \textit{SDO}/HMI line--of--sight photospheric magnetograms showing the magnetic field changes near the inner polarity inversion line (PIL) region (shown by the yellow circles). (d) The field lines obtained from the NLFFF extrapolation method. The low--lying sheared field lines and the high--lying less sheared field lines are shown by the purple and blue colors, respectively. A set of field lines with a relatively higher twist are represented by orange color lines. It is likely that the low lying sheared field lines would form the flux rope due to magnetic field changes at the inner PIL.}
\label{fig3}
\end{figure*}

Detailed structures of the photospheric and overlying coronal magnetic field configuration of AR are shown in Figure~\ref{fig2}. Figure~\ref{fig2}(a) represents the \textit{SDO}/HMI LOS photospheric magnetogram at 04:00:24 UT on 2014 December 17. It can be observed that the AR consists of a compact patch of positive polarity (shown by the green oval) surrounded by negative polarity region (shown by the white oval). This type of photospheric magnetic field configuration is suggestive of a 3D null--point magnetic topology in the corona \citep[see e.g.,][]{Joshi15,Joshi17,Joshi18}. The field lines enclosed below the fan should connect the central compact positive polarity to the surrounding negative polarity. There also exists a bipolar type magnetic field configuration in between the central compact positive polarity and the outer negative polarity with a sharply defined PIL that resembles with an inverted S--letter (see the region enclosed by the red box in Figure~\ref{fig2}(a)). Hereafter, we call this inverted S--shaped PIL as inner PIL. It can be anticipated that the field lines from the outer negative polarity field region (shown by the white oval in Figure~\ref{fig2}(a)) should connect to the far south--westward positive polarity region (shown by the yellow oval Figure~\ref{fig2}(a)).

\begin{figure*}
\vspace*{1cm}
\centerline{
	\hspace*{0\textwidth}
	\includegraphics[width=1.0\textwidth,clip=]{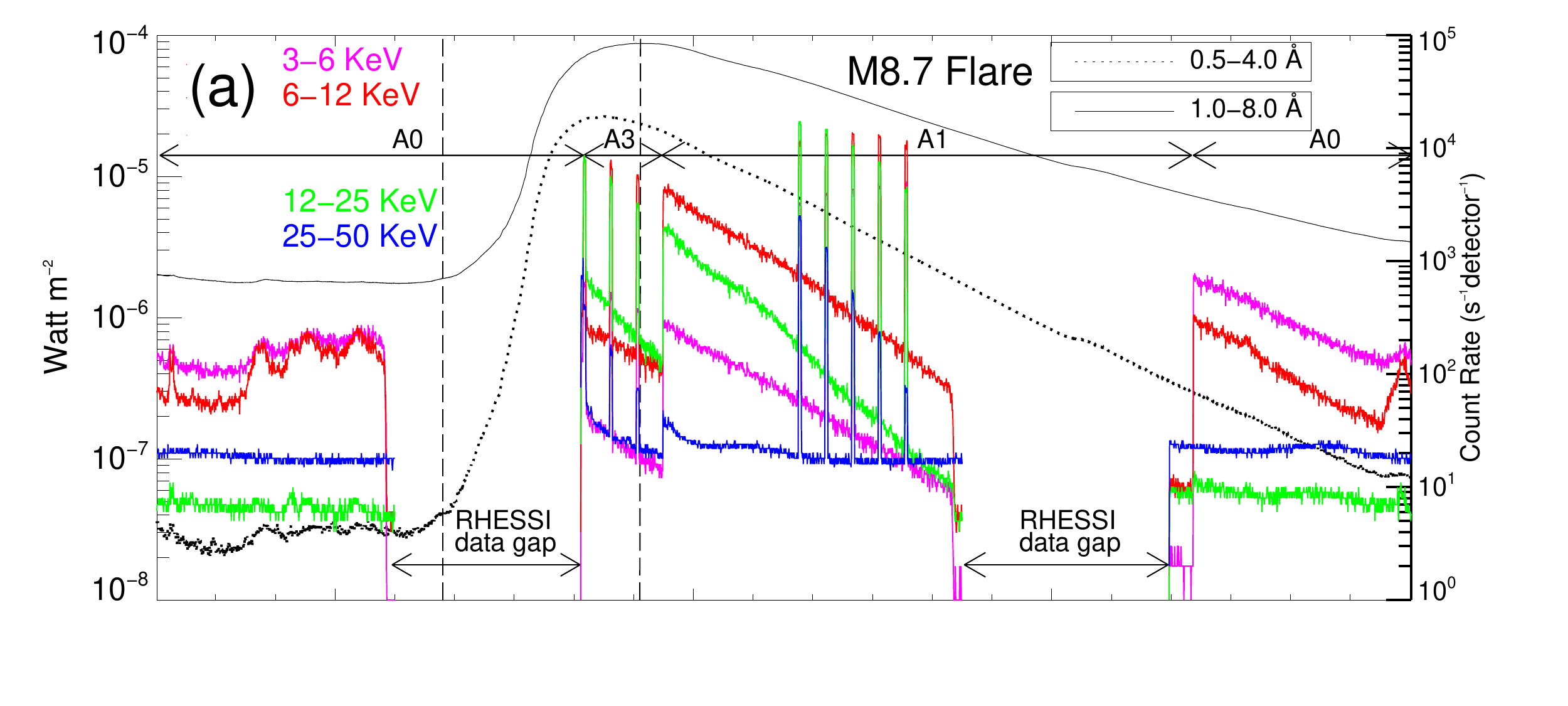}
	}
\vspace*{-1.5cm}
\centerline{
	\hspace*{0\textwidth}
	\includegraphics[width=1.0\textwidth,clip=]{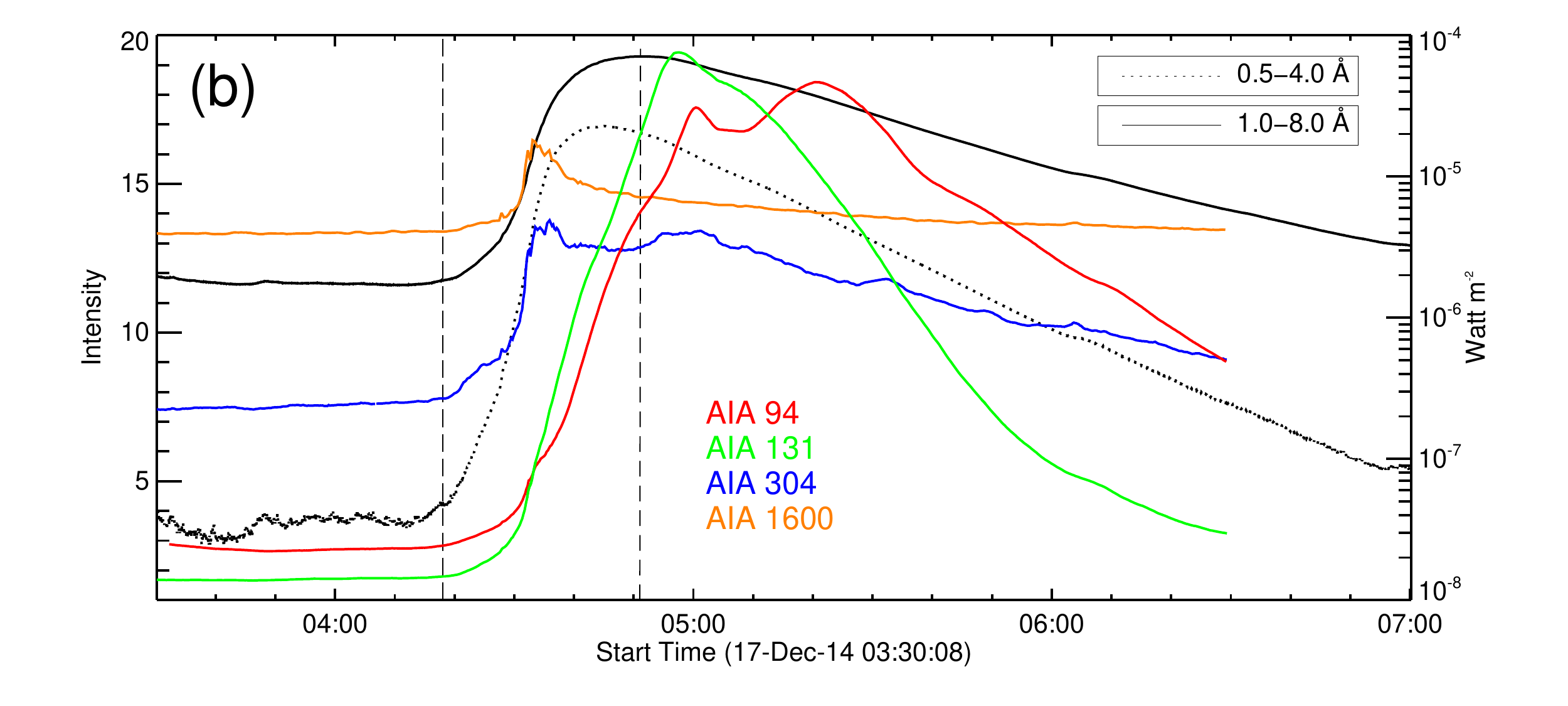}
	}
\vspace*{0cm}
\caption{(a) {\it GOES} and {\it RHESSI} X--ray temporal profiles at different energy channels from $\approx$03:30 UT to $\approx$07:00 UT on 2014 December 17. The black solid and dotted lines represent the GOES X--ray flux profiles at 1.0--8.0 and 0.5-4.0~\AA, respectively. The pink, red, green, and blue lines show the RHESSI X--ray profiles at 3--6, 6--12, 12--25, and 25--50 keV energy bands, respectively. Data gaps between $\approx$04:10 UT--$\approx$04:40 UT and $\approx$05:45 UT--$\approx$06:19 UT are due to the night time and/or South Atlantic Anomaly (SAA) period of the spacecraft. (b) \textit{SDO}/AIA intensity profiles from $\approx$03:30 UT to $\approx$06:30 UT on 2014 December 17. The red, green, blue, and orange curves represent the AIA intensity profiles at 94, 131, 304, and $\rm 1600~\AA,$ respectively. The GOES X--ray profiles at 1.0--8.0 \AA\ (solid black) and 0.5--4.0 \AA\ (dotted black) are overplotted for comparison.}
\label{fig4}
\end{figure*}

In order to formalize the coronal magnetic topology of AR 12242, PFSS and NLFFF extrapolation methods have been used. Both the extrapolation methods show a closed 3D null--point topology over the AR (see PFSS and NLFFF results in Figures~\ref{fig1}(a) and~\ref{fig2}(b)--(d), respectively). The NLFFF analysis reveals that the central positive polarity region is connected with the outer negative polarity regions by a set of closed field lines (shown by the yellow color in Figures~\ref{fig2}(b)--(d)). The outer negative polarity regions are also connected with the far south--westward positive polarity region (i.e., the region co--spatial with remote brightening (RB), described in Section~\ref{sec4}) by another set of closed field lines (shown by red color in Figures~\ref{fig2}(b)--(d)). The null--point lies in a region between the inner yellow and outer red field lines, over the central compact polarity (shown by a cyan isosurface in Figures~\ref{fig2}(b)--(d)). This isosurface represents the magnetic field of 4 Gauss and situated at a height of $\approx$36$\pm$1 Mm (shown by the arrow in Figure~\ref{fig2}(d)).

Magnetic field near the inner PIL (shown inside the red box in Figure~\ref{fig2}(a)) persistently experienced small--scale yet significant changes including converging motions and moving magnetic features. The areas showing these changes are marked by the three yellow circles (see Figures~\ref{fig3}(a)--(c)). In the middle circle, we clearly see the disappearance of few negative polarity patches. Within the right side yellow circle, we see flux emergence as well as the westward--moving magnetic features. We also observed similar small flux changes within the left yellow circle. All these changes can be seen clearly in the \textit{SDO}/HMI animation attached with Figure~\ref{fig3}. These magnetic changes (flux cancellation, emergence, and moving magnetic features) close to the PIL indicate that some magnetic stress and energy are injected in the AR through the photospheric layer. More importantly, the NLFFF extrapolation result demonstrates the presence of an extended set of low--lying sheared field lines along the inner PIL which are shown by the purple--colored lines in Figure~\ref{fig3}(d). We also obtained a few relatively highly sheared field lines which are represented by orange color in Figure~\ref{fig3}(d). The observed changes in the magnetic field at the inner PIL play a crucial role in the formation process of the flux rope via tether--cutting reconnection involving the low--lying sheared arcades (see Subsection~\ref{sec4.2.1} and Figure~\ref{fig5}). The converging motion of the magnetic flux elements toward the PIL and flux cancellation are considered to be the key ingredients of the tether--cutting process \citep[][]{Moore92,Moore01}. Several earlier works confirm the role of these observed processes during the formation and evolution of the flux rope \citep[e.g.,][]{van89,Joshi2017,Yardley18,Joshi20}.


\section{Multi--wavelength observations of the eruptive flare}
\label{sec4}


\subsection{Flux profiles of \textit{GOES}, \textit{RHESSI}, and AIA}
\label{sec4.1}

The temporal evolutions of X--ray flux (from {\it GOES} and {\it RHESSI}) and (E)UV intensity (from AIA) during the flare are shown in Figure~\ref{fig4}. The flare has been classified as an M8.7 class flare in soft X--ray (Figure~\ref{fig4}(a)). The soft X--ray flux in 1.0--8.0 \AA\ and 0.5-4.0~\AA\ wavelengths undergo monotonic increase from $\approx$04:18 UT, implying the onset of the flare. The flare emission peaked at $\approx$04:51 UT in GOES 1.0--8.0 \AA\ channel. It consists of a long decay phase that lasts up to $\approx$07:00 UT. {\it RHESSI} X--ray flux profiles at different energy bands (3--50 keV) are also overplotted in Figure~\ref{fig4}(a). {\it RHESSI} observed this flare event during the pre--flare phase, i.e., from $\approx$03:40 UT to $\approx$04:10 UT, as well as, most of the peak phase from $\approx$04:40 UT to $\approx$05:45 UT. There are two gaps in {\it RHESSI} observations: during the rise and part of the peak phase of the flare from $\approx$04:10 UT to $\approx$04:40 UT as well as from $\approx$05:45 UT to $\approx$06:19 UT during the decay phase.

Figure~\ref{fig4}(b) represents the AIA light curves at four wavelengths, viz. 94, 131, 304 and 1600 \AA. We have estimated the average intensity within the region of interest (shown by the white rectangle in Figure~\ref{fig1}(d)) to get the intensity flux. All intensity profiles show build--up of corresponding fluxes at around $\approx$04:18 UT. The 1600 and 304 \AA\ profiles show the corresponding peaks during the impulsive rise of the GOES soft X--ray flux at $\approx$04:33 UT and $\approx$04:35 UT, respectively. The AIA 131 and 94 \AA\ profiles, on the other hand, show the peak during the decline phase of the GOES soft X--ray light curves at $\approx$05:00 UT, implying hot thermal emission from flare arcades. 

\begin{figure*}
\vspace*{0cm}
\centerline{
	\hspace*{0\textwidth}
	\includegraphics[width=0.95\textwidth,clip=]{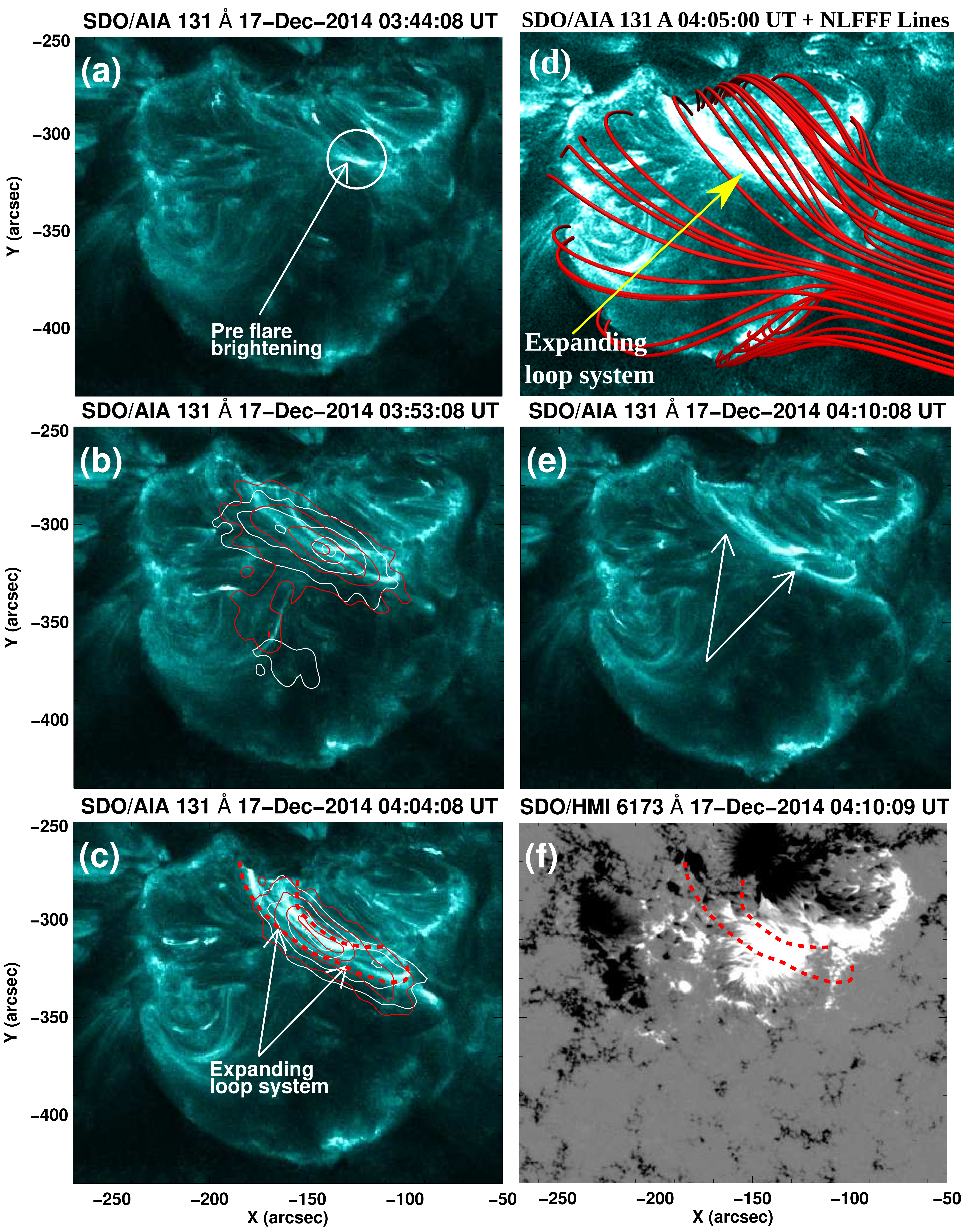}
	}
\vspace*{0cm}
\caption{((a)--(e)) Sequence of selected AIA 131 \AA\ images from 03:53:08 UT to 04:10:08 UT, showing the formation and evolution of the flux rope. The white circles in panel (a) show the region of compact brightening. White and red contours in the panels (b)--(c) are the RHESSI X--ray contours of 3--6 and 6--12 keV energy bands, respectively. The contours levels are 45\%, 60\%, 80\%, and 95\% of the peak intensity. (d) \textit{SDO}/AIA 131 \AA\ images at $\approx$04:05 UT with NLFFF extrapolated field lines. (f) \textit{SDO}/HMI photospheric magnetogram at 04:10:09 UT on 2014 December 17. 
The red dashed curved lines in panels (c) and (f) outline the expanding twisted loop system at 04:04:08 which presumably represents a flux rope.}
\label{fig5}
\end{figure*}

\subsection{Dynamical evolution of the flux rope, associated flare and CME}
\label{sec4.2}


\subsubsection{Pre-flare processes and possible scenario of flux rope formation}
\label{sec4.2.1}

\begin{figure*}
\vspace*{1cm}
\centerline{
	\hspace*{0\textwidth}
	\includegraphics[width=1\textwidth,clip=]{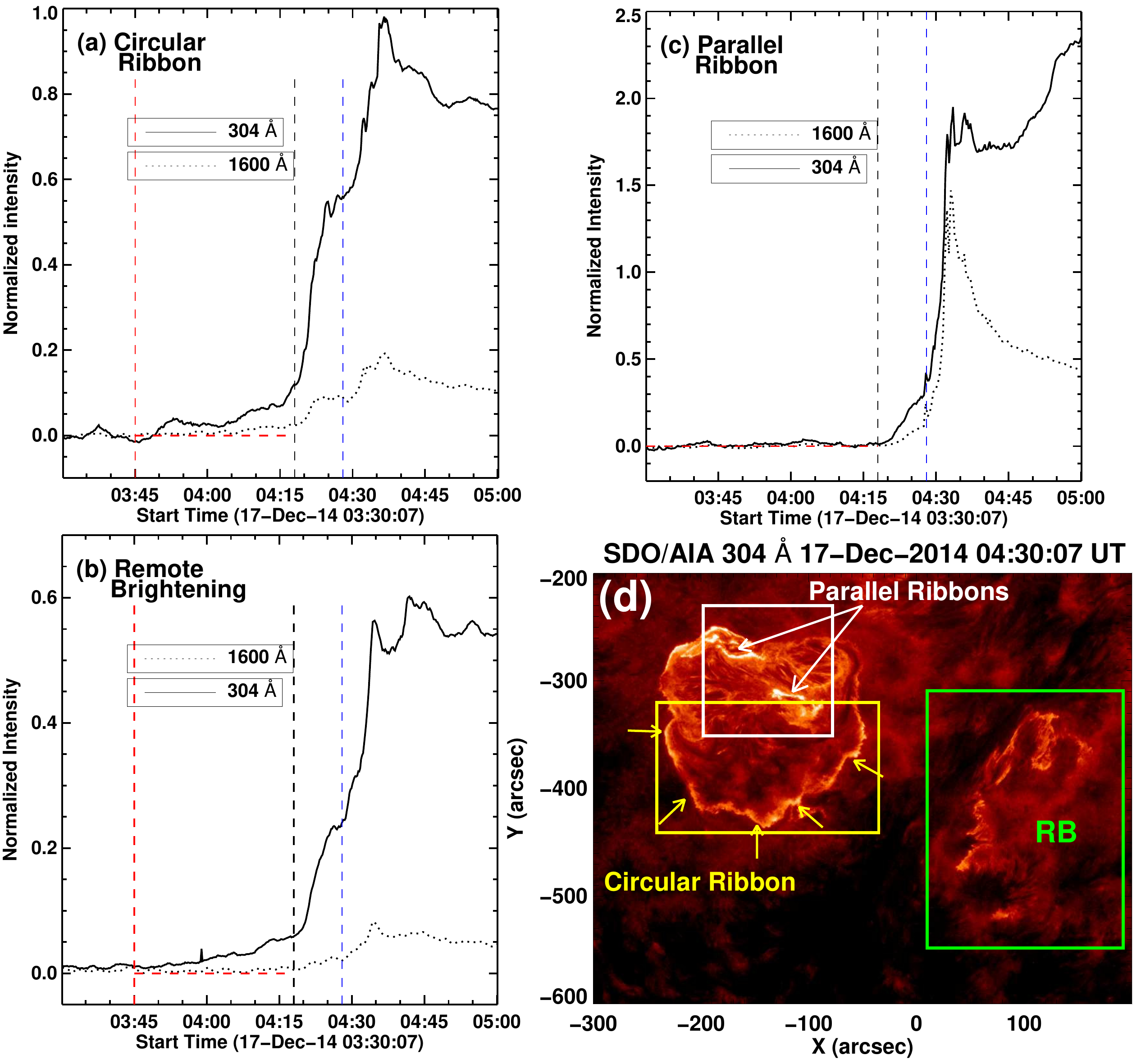}
	}
\vspace*{0cm}
\caption{Normalized intensity profiles of the circular ribbon (a), remote brightening region (b) and parallel ribbons (c), respectively at 304 (solid line) and 1600 (dotted line) \AA\ wavelength channels. The red and black vertical dashed lines represent the start ($\approx$03:45 UT) and the end time ($\approx$04:18 UT) time of the pre--flare phase, respectively. The verticle blue line represent the start time of another phase at $\approx$04:28 UT. (d) \textit{SDO}/AIA 304 \AA\ image at 04:30:07 UT on 2014 December 17. Yellow, white and green boxes represent the area used for intensity measurements for circular ribbon, parallel ribbons, and remote brightening region, respectively.}
\label{fig6}
\end{figure*}

Pre--eruptive phase starts at $\approx$03:40 UT and lasts till $\approx$04:15 UT. Figure~\ref{fig5} shows the selected images during this phase at AIA 131 \AA\ wavelength channel. The compact brightening appears to start at $\approx$03:44 UT near the inner PIL (see the area shown by the red box in Figure~\ref{fig2}(a)). This compact brightening region is shown in Figure~\ref{fig5}(a) by white circle.
Along with this compact brightening, we also see the appearance of loop--like structures (shown by the arrows in Figures~\ref{fig5}(b)--(e)) connecting the westward positive polarity to the eastward negative polarity regions (cf. dotted curved lines in Figures~\ref{fig5}(c) and~\ref{fig5}(f)). The successive formation of these channel--like twisted loop structures and corresponding footpoint brightenings, provide indirect evidence for the build--up of a flux rope, probably due to the tether--cutting reconnection (see also the animations of AIA 131 and 94 \AA\ associated with Figure~\ref{fig5}). It is also important to note that the formation process of flux rope occurs inside the fan--dome (see Figure~\ref{fig5}(d)). However, from AIA images we could not gather any robust observational structure in support of a flux rope (such as sigmoid, filament or mini-filament) inside the fan-dome during the pre-flare phase. On the other hand, as discussed in Section~\ref{sec4.2.4}, the eruption began from underneath the dome during the flare's impulsive phase which supports the idea of in-situ formation of the flux rope. 

\begin{figure*}
\vspace*{-4.5cm}
\centerline{
	\hspace*{0.05\textwidth}
	\includegraphics[width=0.65\textwidth,clip=]{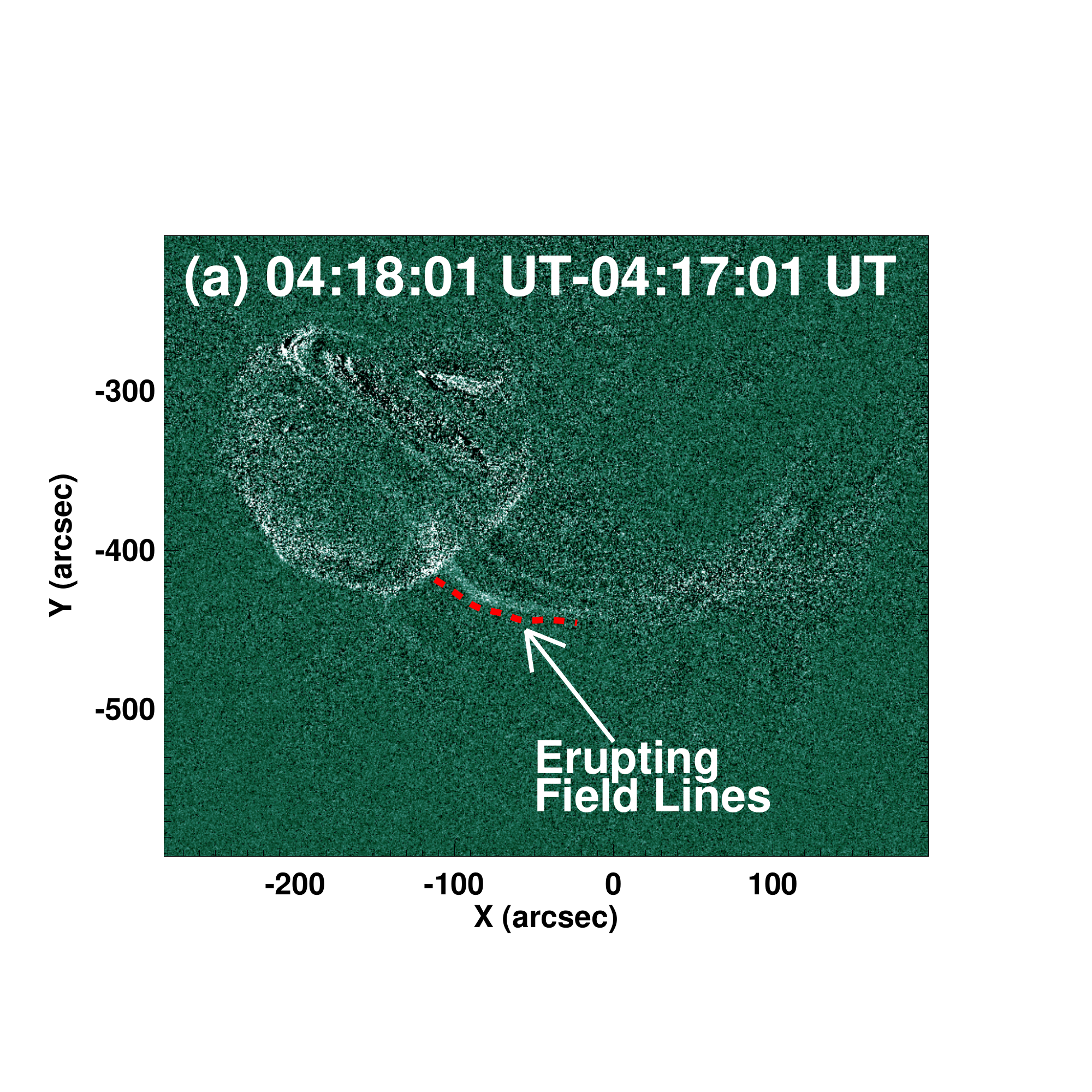}
	\hspace*{-0.255\textwidth}
	\includegraphics[width=0.65\textwidth,clip=]{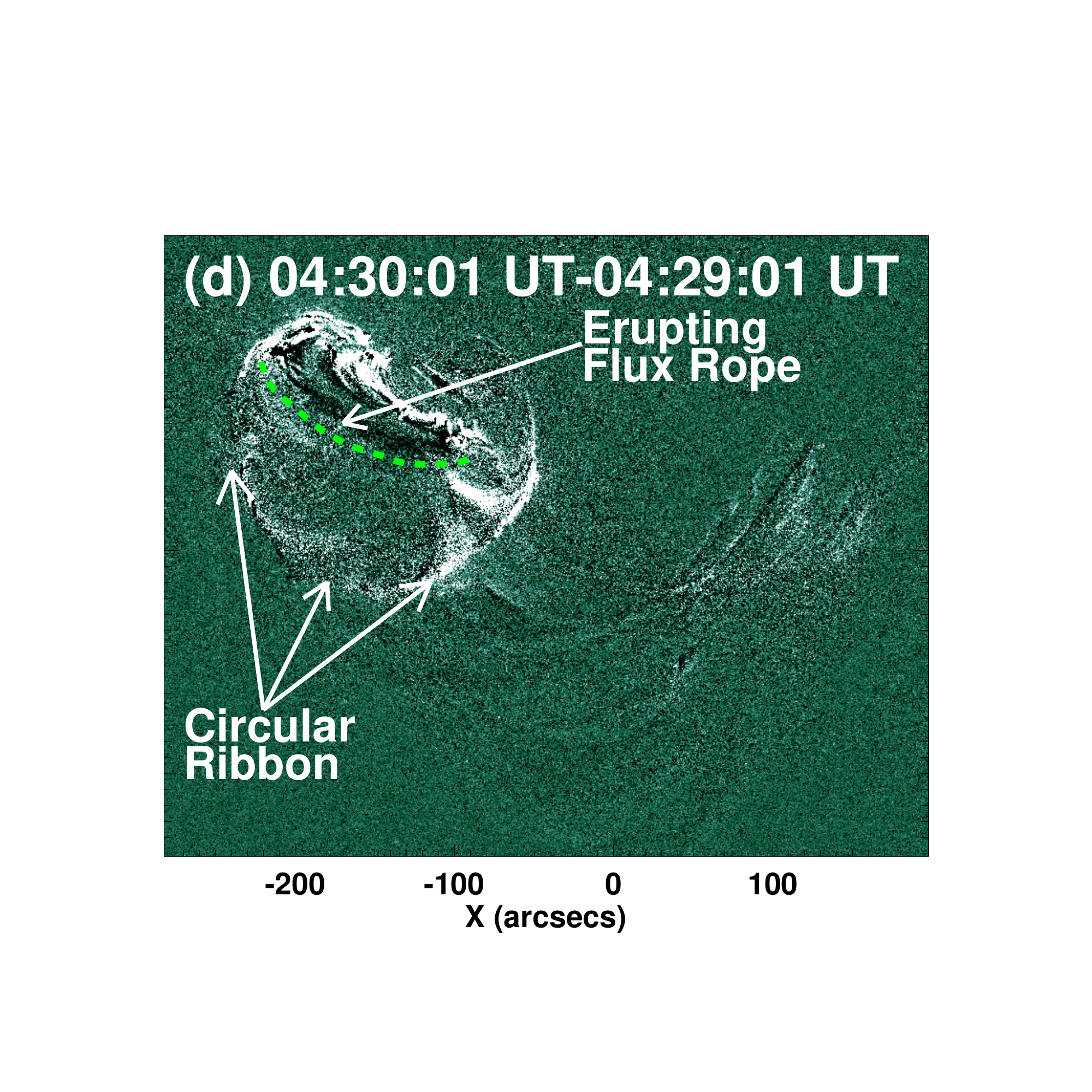}
	}
\vspace*{-5.73cm}
\centerline{
	\hspace*{0.05\textwidth}
	\includegraphics[width=0.65\textwidth,clip=]{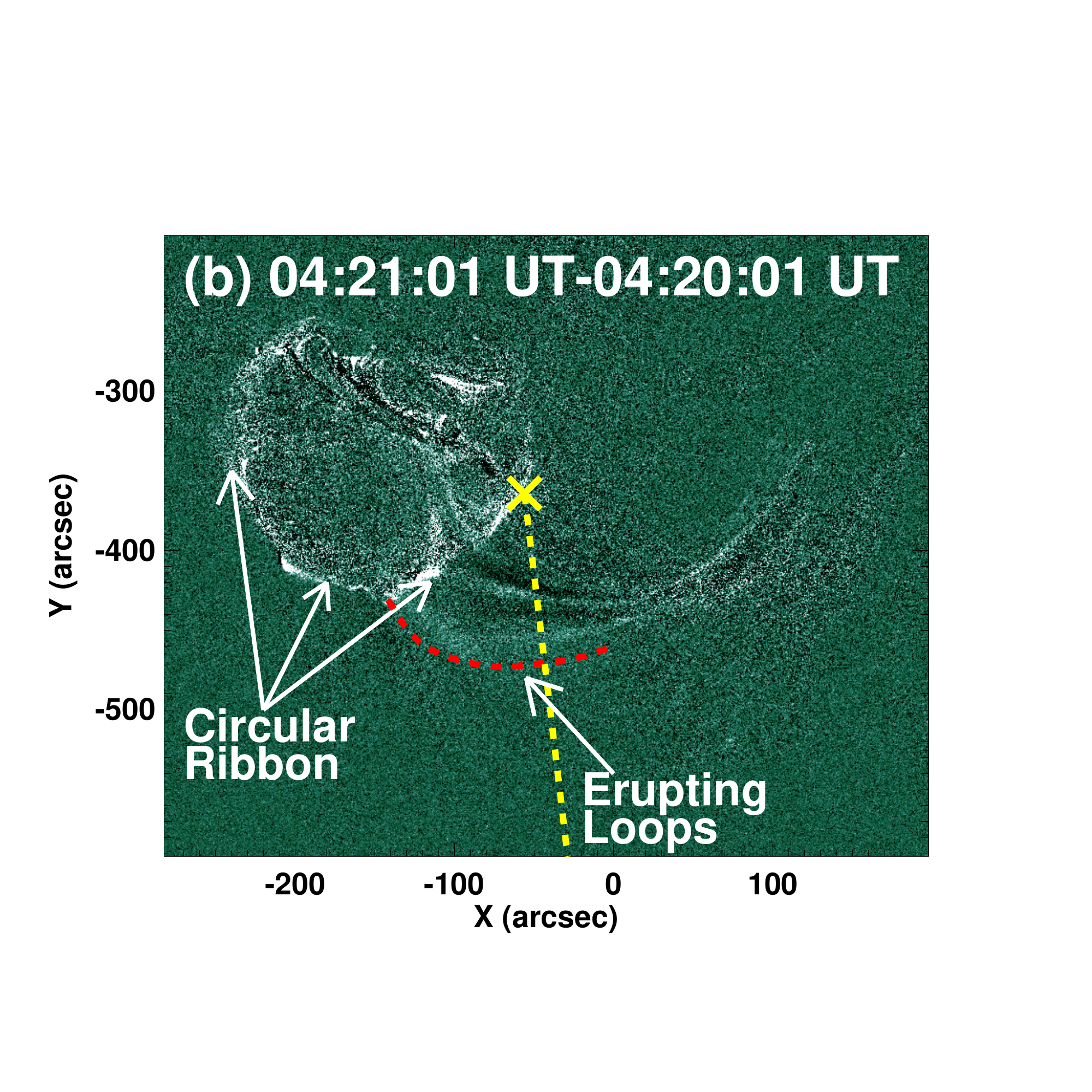}
	\hspace*{-0.255\textwidth}
	\includegraphics[width=0.65\textwidth,clip=]{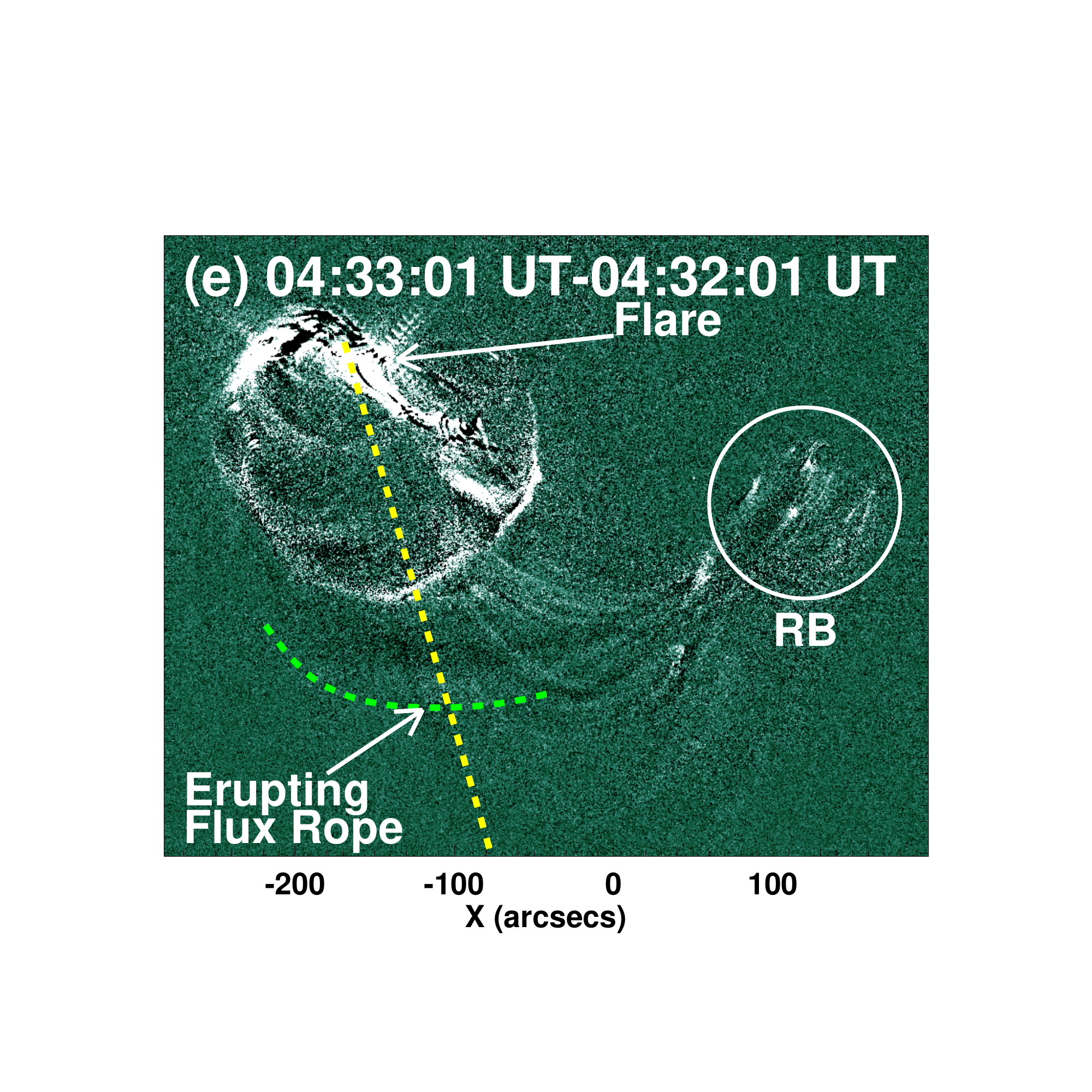}
	}
\vspace*{-5.73cm}
\centerline{
	\hspace*{0.05\textwidth}
	\includegraphics[width=0.65\textwidth,clip=]{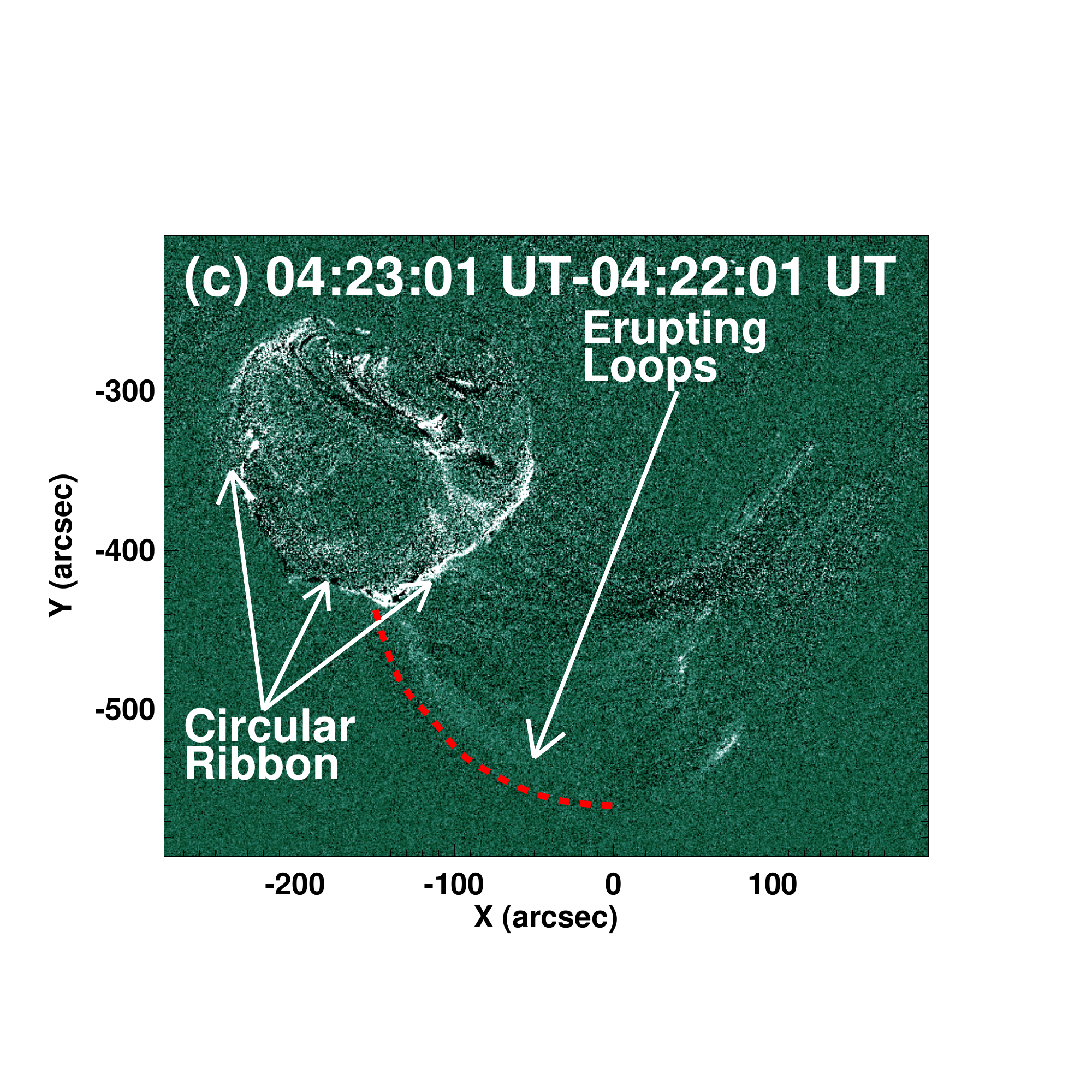}
	\hspace*{-0.255\textwidth}
	\includegraphics[width=0.65\textwidth,clip=]{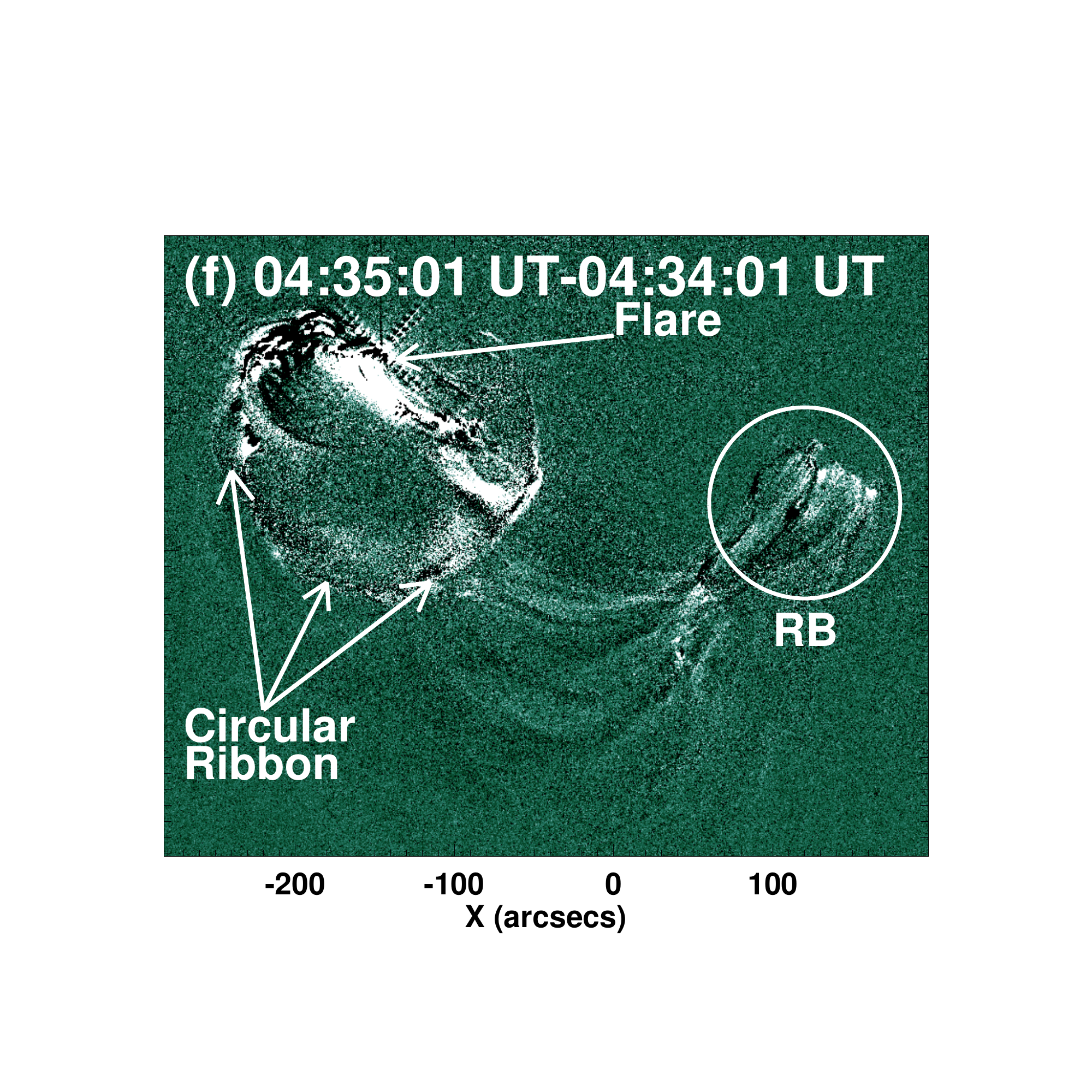}
	}
\vspace*{-1.5cm}
\caption{((a)--(f)) \textit{SDO}/AIA 94 \AA\ running difference images showing the eruptions during the event. The red and green dashed curves represent the eruption of the erupting loops and the flux rope, respectively. The white circle in panels (e) and (f) shows the remote brightening region. The vertical yellow dashed lines in panels (b) and (e) represent the chosen trajectories along which the height measurements for eruptions are made (see height--time plot in Figure~\ref{fig10}(a)).}
\label{fig7}
\end{figure*}

\subsubsection{Temporal evolution of circular and parallel ribbons}
\label{sec4.2.2}

A careful examination of AIA 304 \AA\ image sequence clearly shows that the circular ribbon formation started well before the impulsive phase of the flare while the development of parallel ribbons initiated during the impulsive phase.  In order to critically distinguish the timing of the brightenings in the circular ribbon and the parallel ribbons, we precisely compute the intensity profiles of the flare ribbons which are presented in Figure~\ref{fig6}(a)--\ref{fig6}(c). To plot these profiles, we have estimated the intensity of the respective regions (see Figure~\ref{fig6}(d)) from $\approx$03:30 UT onward using series of AIA 304 and 1600 \AA\ images (see solid and dotted lines for the two channels, respectively). To estimate the normalized intensity, we first subtract the background intensity from the intensity of the selected region and then divide the resulting value by the background intensity again. Here, the background intensity is calculated by taking average of all the intensity values of the images from $\approx$03:30 UT to $\approx$03:45 UT, much before the initiation of the pre--flare phase during which the intensity was observed to be almost constant. 

Time profile of the circular flare ribbon (Figure~\ref{fig6}(a)) readily reveals that its formation started as early as $\approx$03:45 UT and the emission from the circular ribbon gradually rose till 04:18 UT (indicated by the vertical red dashed line). The time profile of the remote ribbon (Figure~\ref{fig6}(b)) reveals that its temporal evolution to be very similar to the circular ribbon: both slowly evolve during $\approx$03:45--04:18 UT, before entering into the phase of impulsive brightening, which is expected. On the other hand, the parallel ribbons developed about more than 30 minutes later, after $\approx$04:18 UT, with the onset of the impulsive phase (Figure~\ref{fig6}(c)). The distinct and simultaneous appearance of the circular ribbon along with the remote ribbon prior to the flare's impulsive phase is indicative of the null--point reconnection.

\begin{figure*}
\vspace*{0cm}
\centerline{
	\hspace*{0\textwidth}
	\includegraphics[width=1.0\textwidth,clip=]{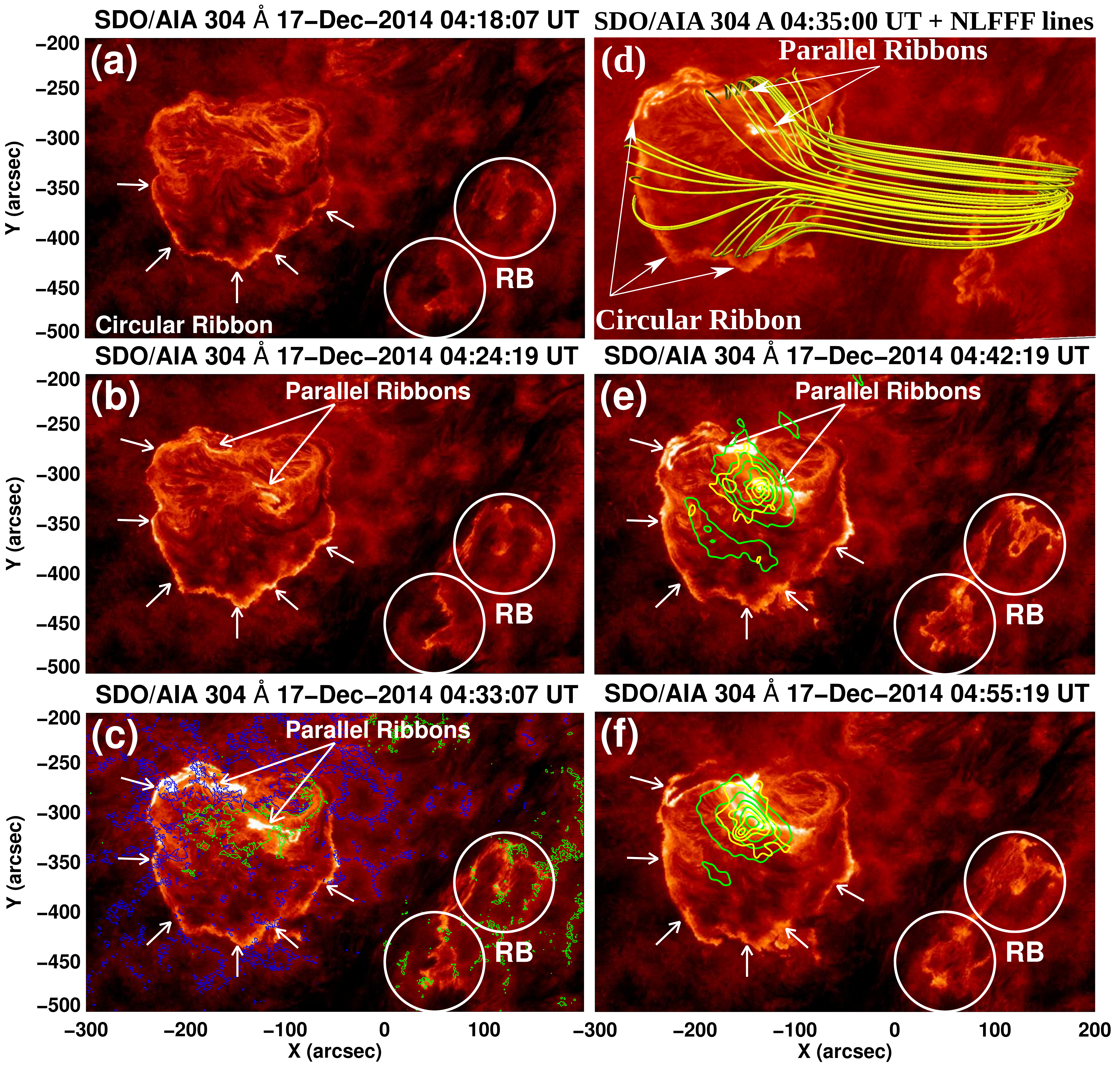}
	}
\vspace*{0cm}
\caption{((a)--(f)) Sequence of \textit{SDO}/AIA 304 \AA\ images showing the evolution of different sets of flare ribbons. The big white circles in the panels (a) to (f) show the remote brightening (RB) regions. The blue and green contours in panel (c) are the \textit{SDO}/HMI magnetogram contours representing the negative and positive polarity regions, respectively. The contour levels are $\pm$150 Gauss. (d) \textit{SDO}/AIA 304 \AA\ image  at $\sim$04:35 UT overplotted with the NLFFF extrapolated field lines. The green and yellow contours in panels (e) and (f) represent the \textit{RHESSI} X--ray contours at 10--15 and 25--50 keV energy bands. Contour levels are 20\%, 40\%, 60\%, 80\%, and 95\% of the peak intensity for 10--15 keV sources and 35\%, 60\%, 80\%, and 95\% for the 25--50 keV sources.}
\label{fig8}
\end{figure*}

\subsubsection{Field line restructuring and eruption}
\label{sec4.2.3}

\begin{figure*}
\vspace*{-3cm}
\centerline{
	\hspace*{0.07\textwidth}
	\includegraphics[width=0.65\textwidth,clip=]{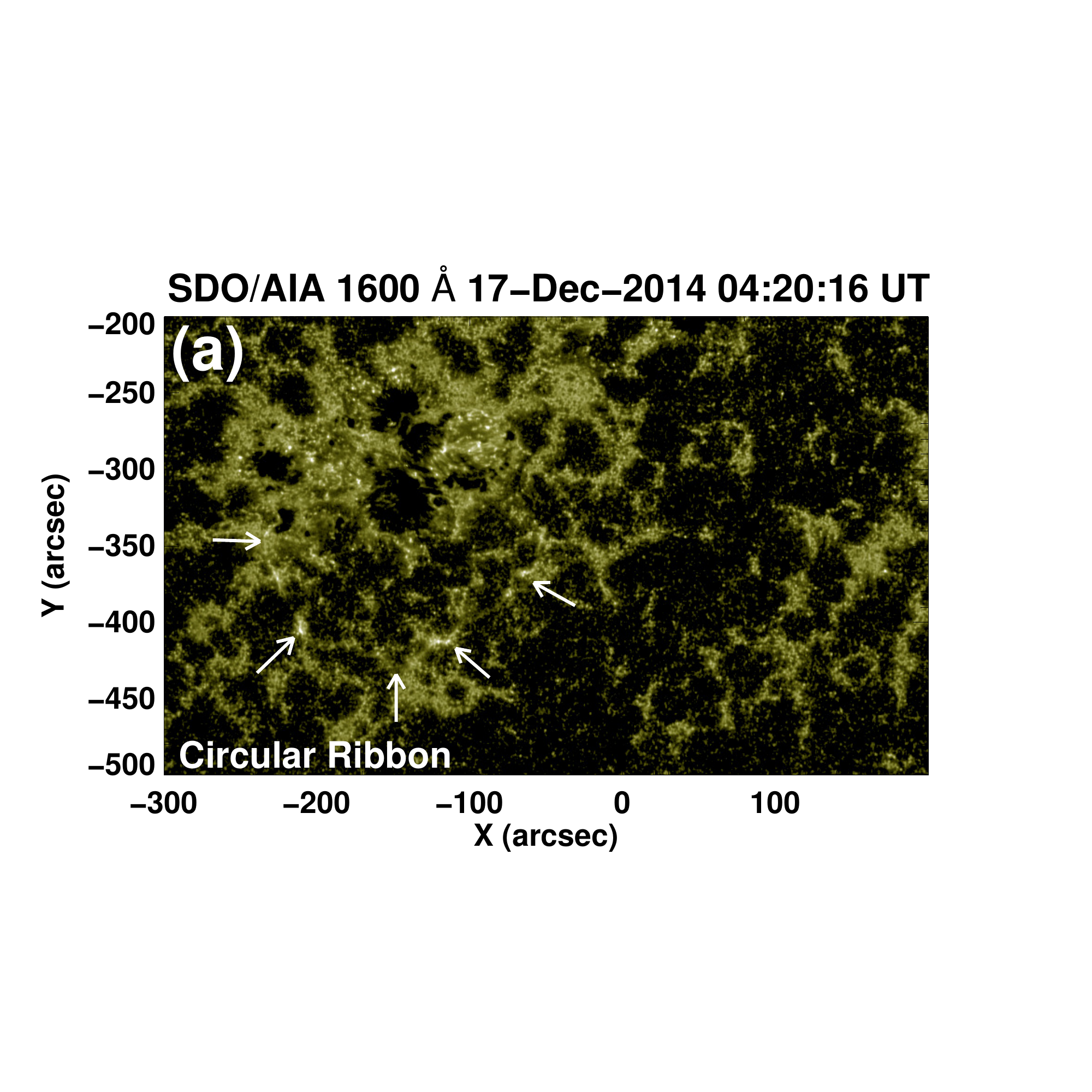}
	\hspace*{-0.22\textwidth}
	\includegraphics[width=0.65\textwidth,clip=]{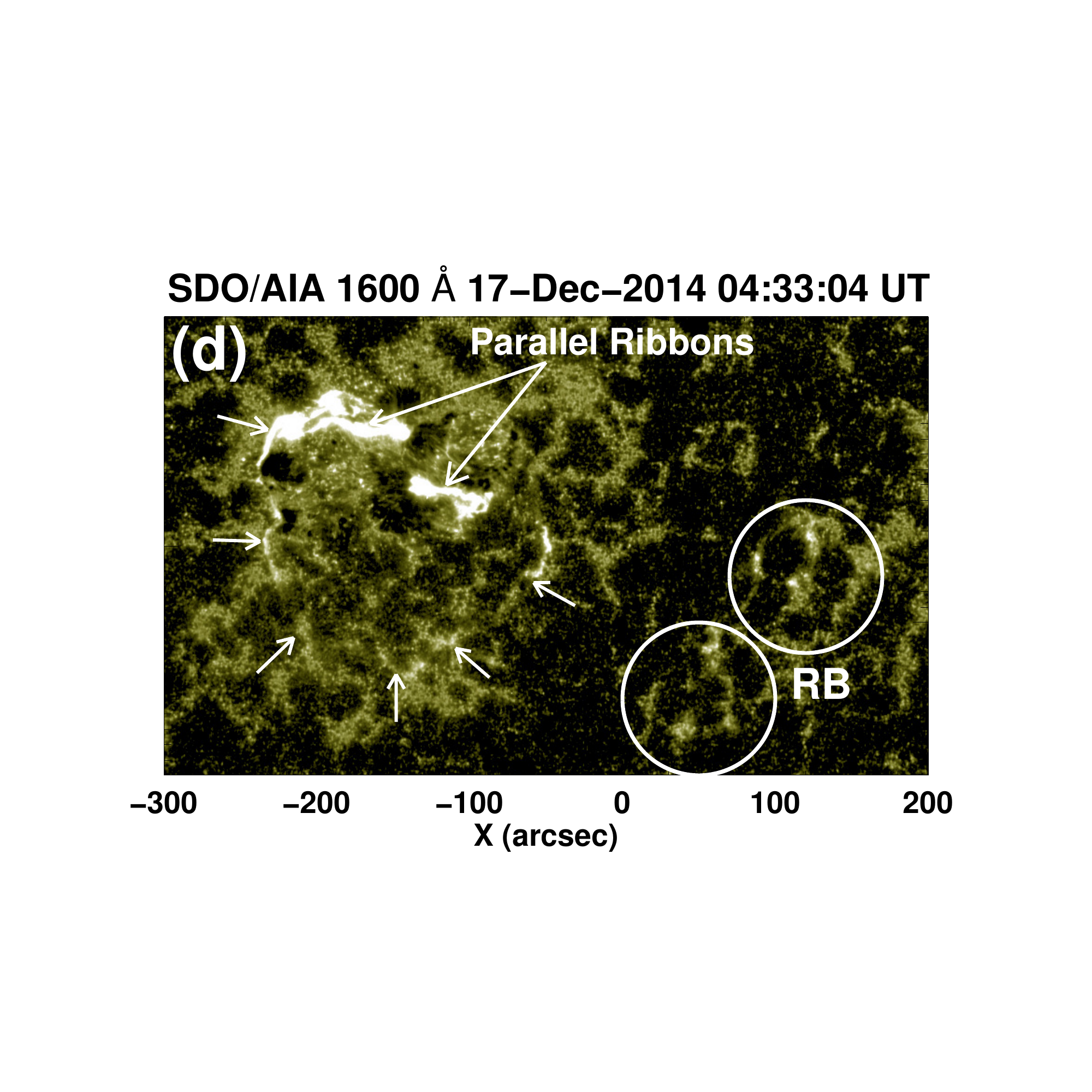}
	}
\vspace*{-6.2cm}
\centerline{
	\hspace*{0.07\textwidth}
	\includegraphics[width=0.65\textwidth,clip=]{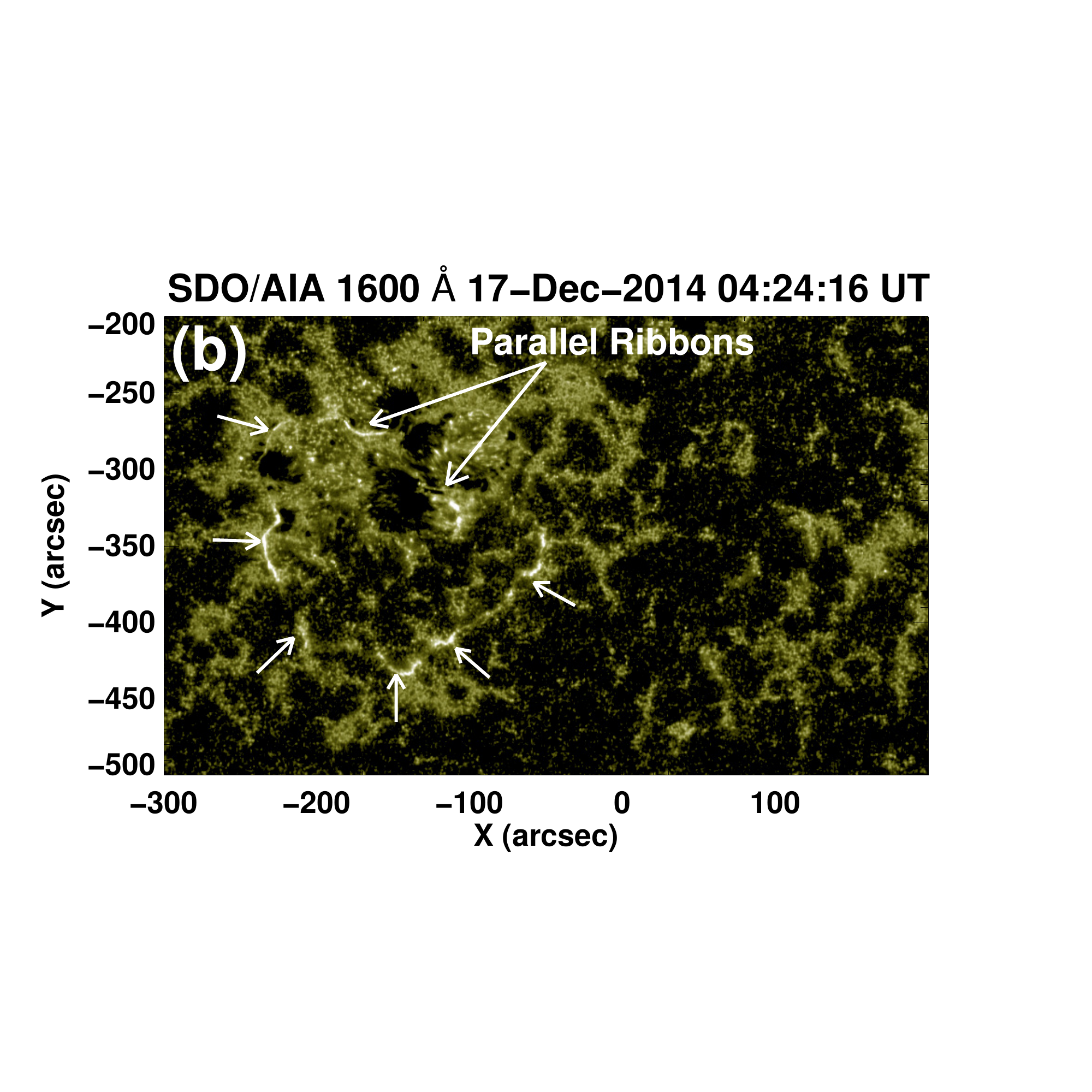}
	\hspace*{-0.22\textwidth}
	\includegraphics[width=0.65\textwidth,clip=]{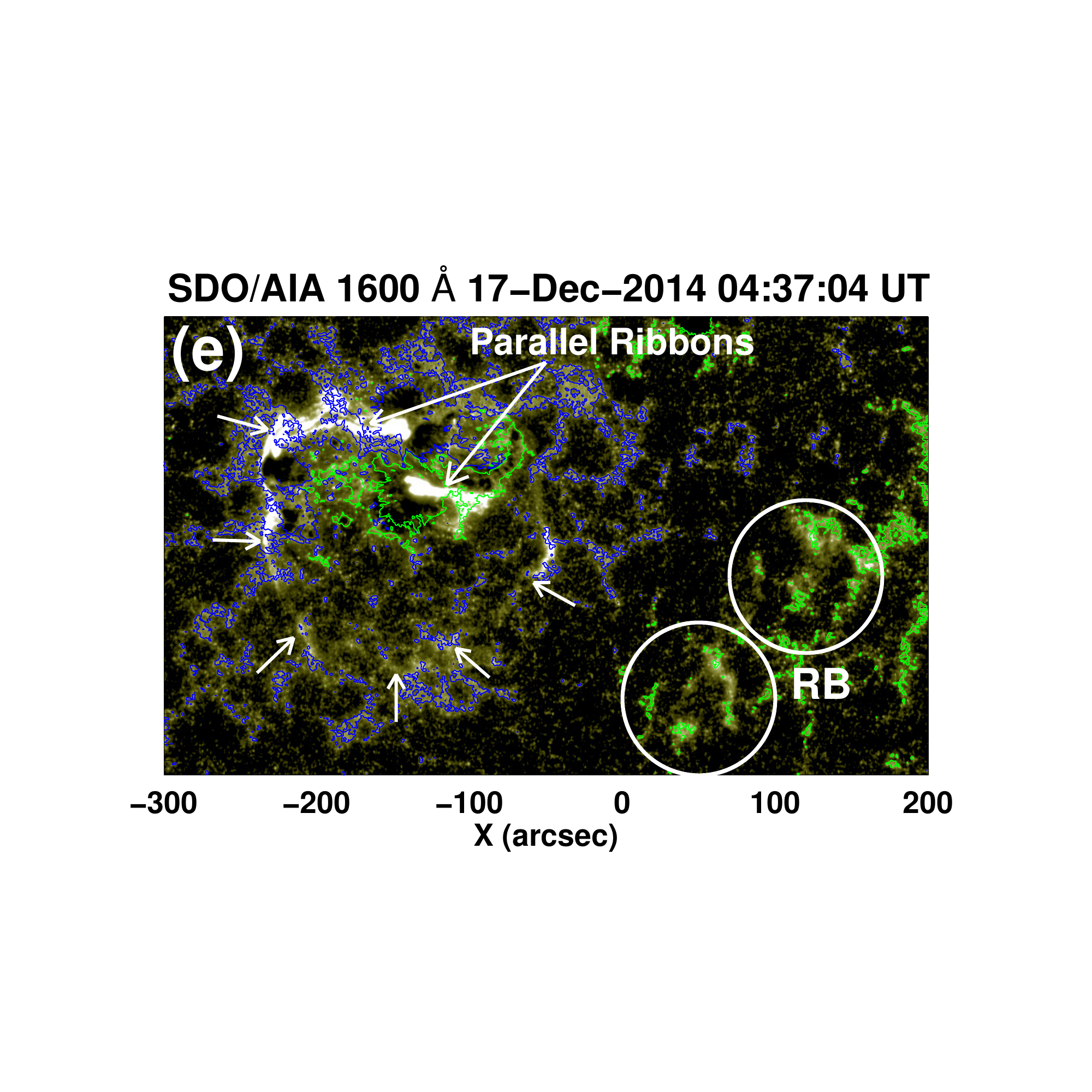}
	}
\vspace*{-6.2cm}
\centerline{
	\hspace*{0.07\textwidth}
	\includegraphics[width=0.65\textwidth,clip=]{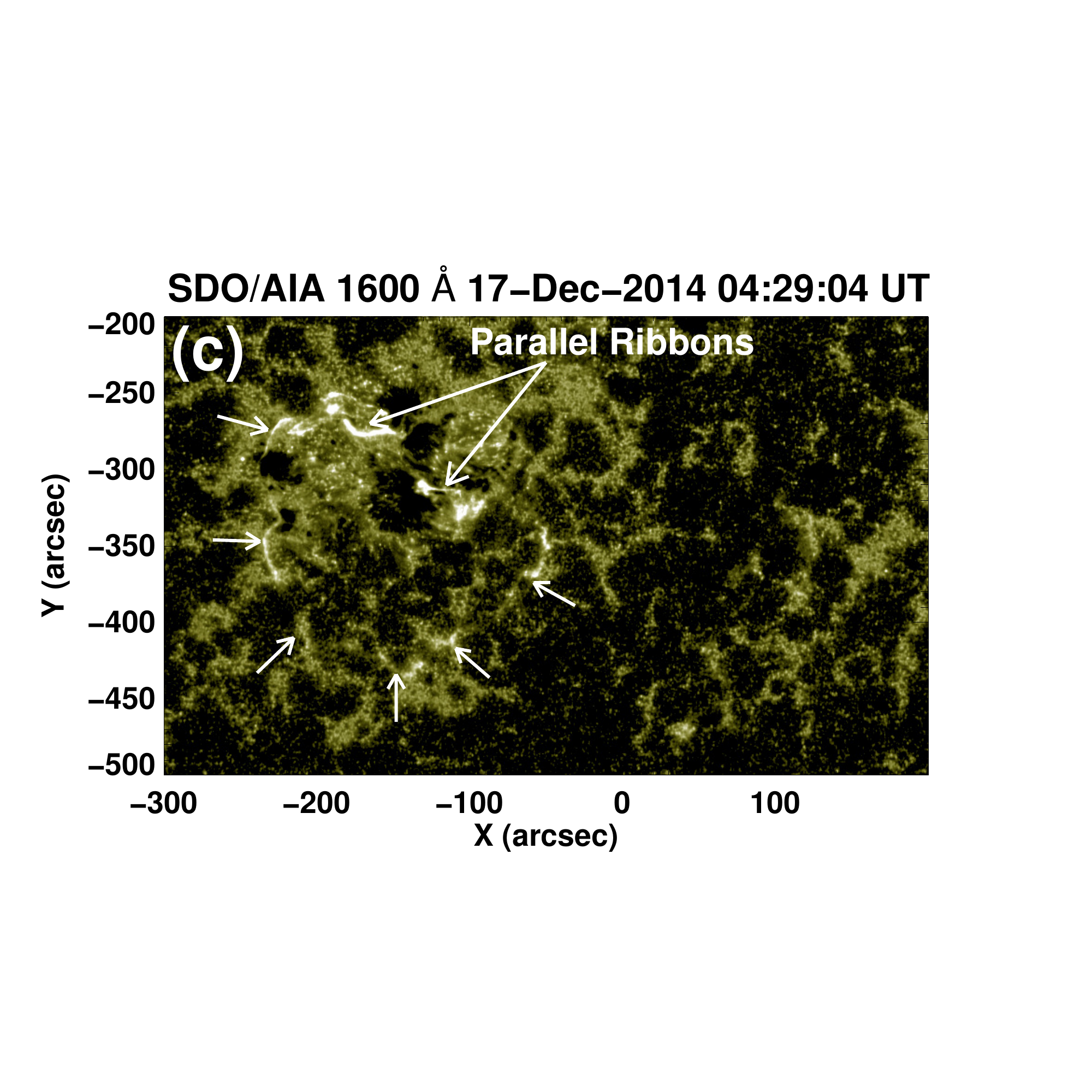}
	\hspace*{-0.22\textwidth}
	\includegraphics[width=0.65\textwidth,clip=]{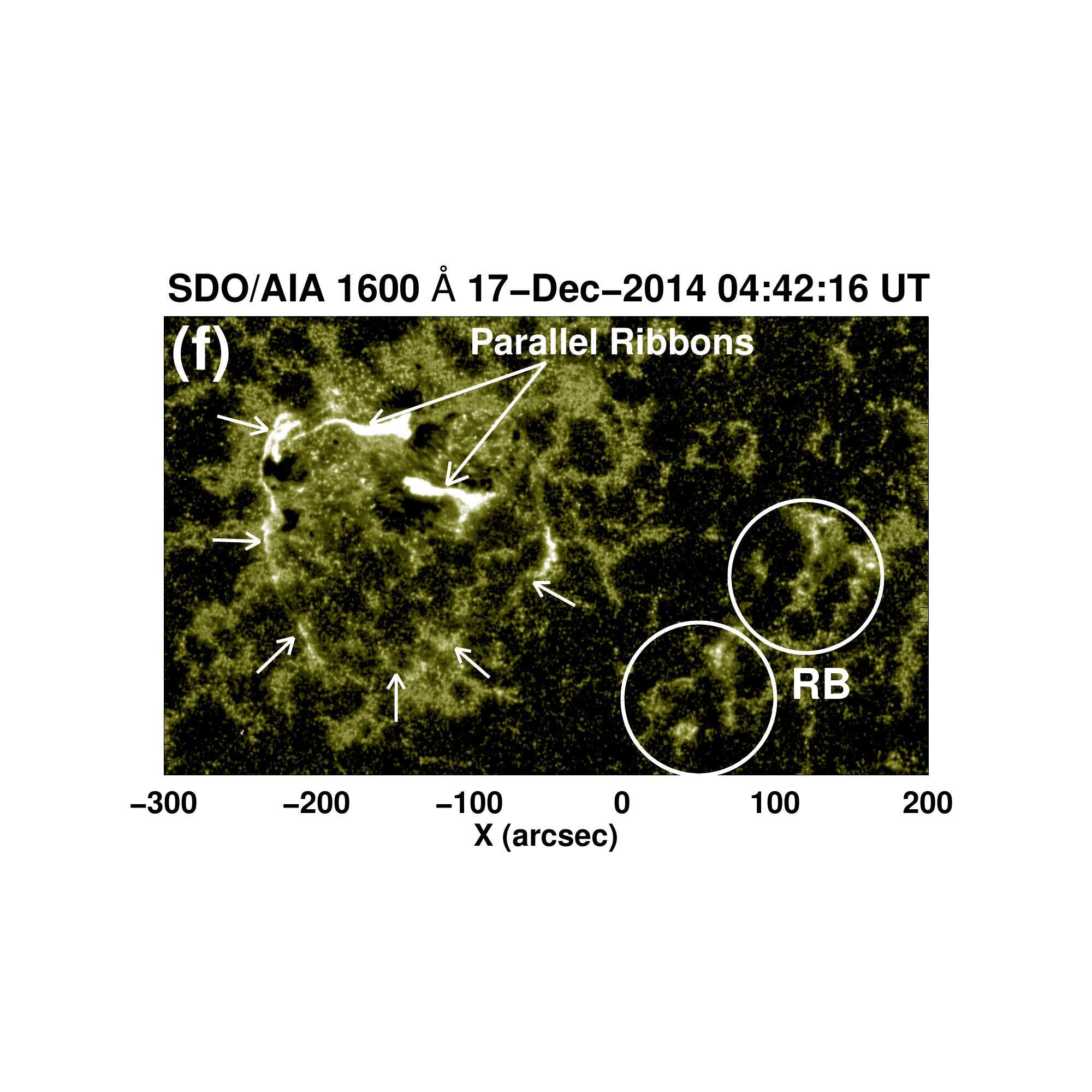}
	}
\vspace*{-2.3cm}
\caption{((a)--(f)) Sequence of \textit{SDO}/AIA 1600 \AA\ images showing the evolution of different sets of flare ribbons. The blue and green contours in panel (e) are the \textit{SDO}/HMI magnetogram contours representing the negative and positive polarity regions, respectively. The contour levels are $\pm$150 Gauss.}
\label{fig9}
\end{figure*}

The flare undergoes a transition into the impulsive phase starting from $\approx$04:18 UT with the outward expansion of hot erupting loops (Figure~\ref{fig7}) and rapid rise in the intensity of circular and parallel ribbons (see Figures~\ref{fig8} and~\ref{fig9}). The erupting loops are observed in AIA 94 \AA\ images which means that these are hot loops ($\sim$6 MK). However, we note that the erupting loops is a visual fact which has been observed in the running difference images of AIA 94 \AA\ (see Figures~\ref{fig7}(a)--(c) and associated animation). This phase is characterized by the onset of restructuring of magnetic field lines over and under the fan structure involving the closed 3D null--point magnetic configuration. The erupting loops are observed until $\approx$04:25 UT (shown by arrows in Figures~\ref{fig7}(a)--(c)). These eruptive loops appear to move southward against the on--disk projection (see red dotted curved lines in Figures~\ref{fig7}(a)--(c)). The erupting loops are more likely those field lines that reconnect continuously in the QSL--halo surrounding the null--point. After reconnection at the magnetic null, some of the accelerated electrons and particles would travel downward along the fan--dome lines, thereby heating the chromospheric region and forming the circular ribbon structure. The accelerated electrons traveling along the field lines surrounding the outer spine would produce RB at their footpoints.

\begin{figure}
\vspace*{-0.3cm}
\centerline{
	\hspace*{0\textwidth}
	\includegraphics[width=1\columnwidth,clip=]{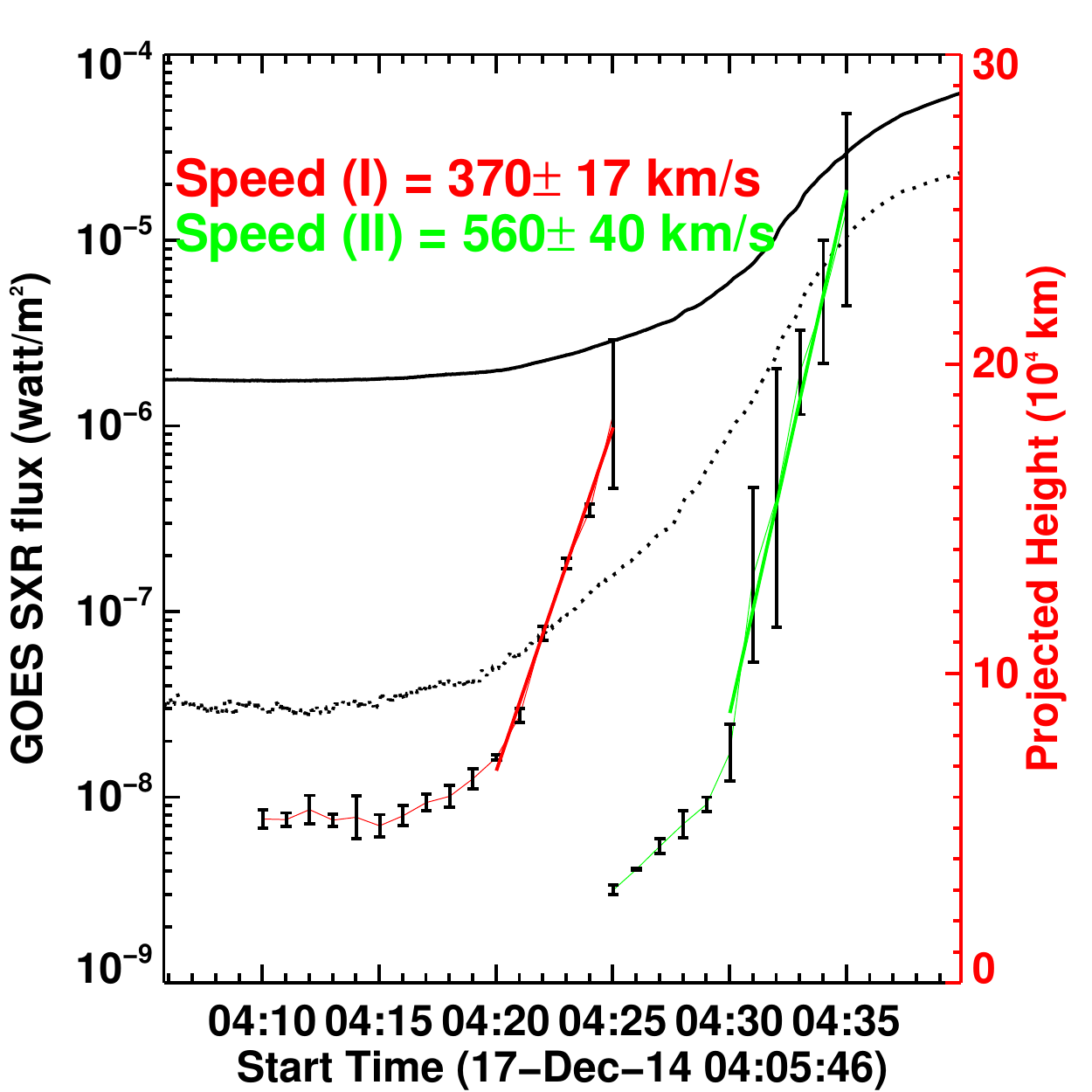}
	}
\vspace*{0cm}
\caption{Projected height--time plots of the erupting loops (I) and flux rope (II) eruptions are shown by the red and green lines, respectively. The GOES soft X--ray flux profiles at 1.0--8.0 \AA\ (solid black line) and 0.5--4.0 \AA\ (dotted black line) are also over--plotted for the comparison. The trajectories along which the height measurements for the erupting coronal loops and flux rope eruptions have been made are shown in Figures~\ref{fig7}(b) and~\ref{fig7}(e) by yellow dashed lines, respectively. Since the erupting features were very faint, we prefer the conventional height--time plot based on the eye judgement over the time--slice diagram.}
\label{fig10}
\end{figure}

The height--time profile showing the eruption of overlying loops, observed in the AIA 94 \AA\ image (Figures~\ref{fig7}(a)--(c)), is shown by the red curve in Figure~\ref{fig10}. It is observed that the eruption of restructured and overlying loops started at $\approx$04:18 UT. The projected height of the erupting structure is estimated along the trajectory shown in Figure~\ref{fig7}(b) by the yellow dashed line. For the purpose, we carefully trace the front of the erupting leading edge along the chosen trajectory. The starting point of the trajectory is used as the reference point (shown by the cross symbol on the yellow dashed line in Figure~\ref{fig7}(b)) for height estimation. We have made three repeated measurements of the height to estimate the uncertainty in the height estimation. The linear speed of the erupting structure comes out to be $ \approx$370 $\rm km~s^{-1}$.

We would like to emphasize that the onset of the eruption of overlying loops temporally matches with the significant and progressive enhancement in the brightening of the parallel, circular and remote ribbons at $\approx$04:18 UT, as evident from the \textit{SDO}/AIA 304 and 1600 \AA\ images (see Figures~\ref{fig8}(a)--(b),~\ref{fig9}(a)--(b),~\ref{fig10}, and associated animations). Furthermore, this stage marks the onset of flare's impulsive phase as recorded in GOES X--ray measurements (see Figure~\ref{fig4}).

\subsubsection{Flux rope eruption and CME initiation}
\label{sec4.2.4}

The appearance of parallel ribbons and the flux rope eruption are the most prominent features of the impulsive phase of the flare (starting from 04:18 UT), which imply the start of the standard flare reconnection. The AIA imaging at 304 \AA\ (Figure~\ref{fig8}) along with corresponding light curve (Figure~\ref{fig6}(c)) show an impulsive rise in the brightness of the parallel ribbons from $\approx$04:18 UT and remain brightened until the late decay phase. The appearance of the parallel ribbons at $\approx$04:18 UT suggest that the flux rope eruption initiated at that time. The AIA 94 \AA\ images clearly show the eruption of hot EUV channel (i.e., flux rope) whose motion can be deduced from the running difference images between $\approx$04:25 UT and $\approx$04:35 UT (see dashed green line in Figures~\ref{fig7}(d)--(f)). The height--time plot of the flux rope (shown by the green plot in Figure~\ref{fig10}) along a chosen trajectory (indicated by the yellow dashed line in Figure~\ref{fig7}(e)) gives the speed of the flux rope to be $\approx$560 $\rm km~s^{-1}$.

\begin{figure}
\vspace*{-2.5cm}
\centerline{
	\hspace*{0\textwidth}
	\includegraphics[width=0.852\textwidth,clip=]{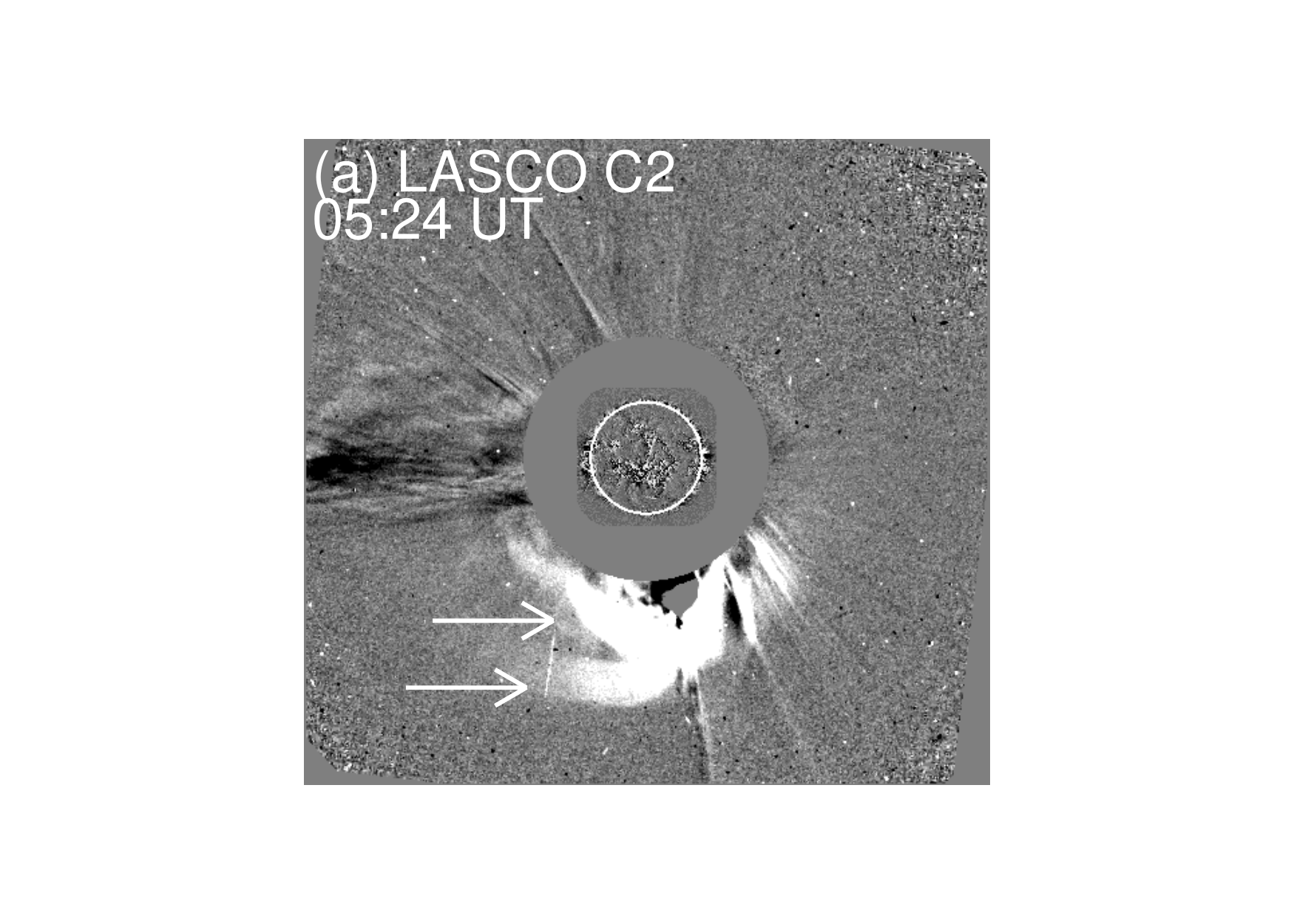}
	}
\vspace*{-3.3cm}
\centerline{
	\hspace*{0\textwidth}
	\includegraphics[width=0.852\textwidth,clip=]{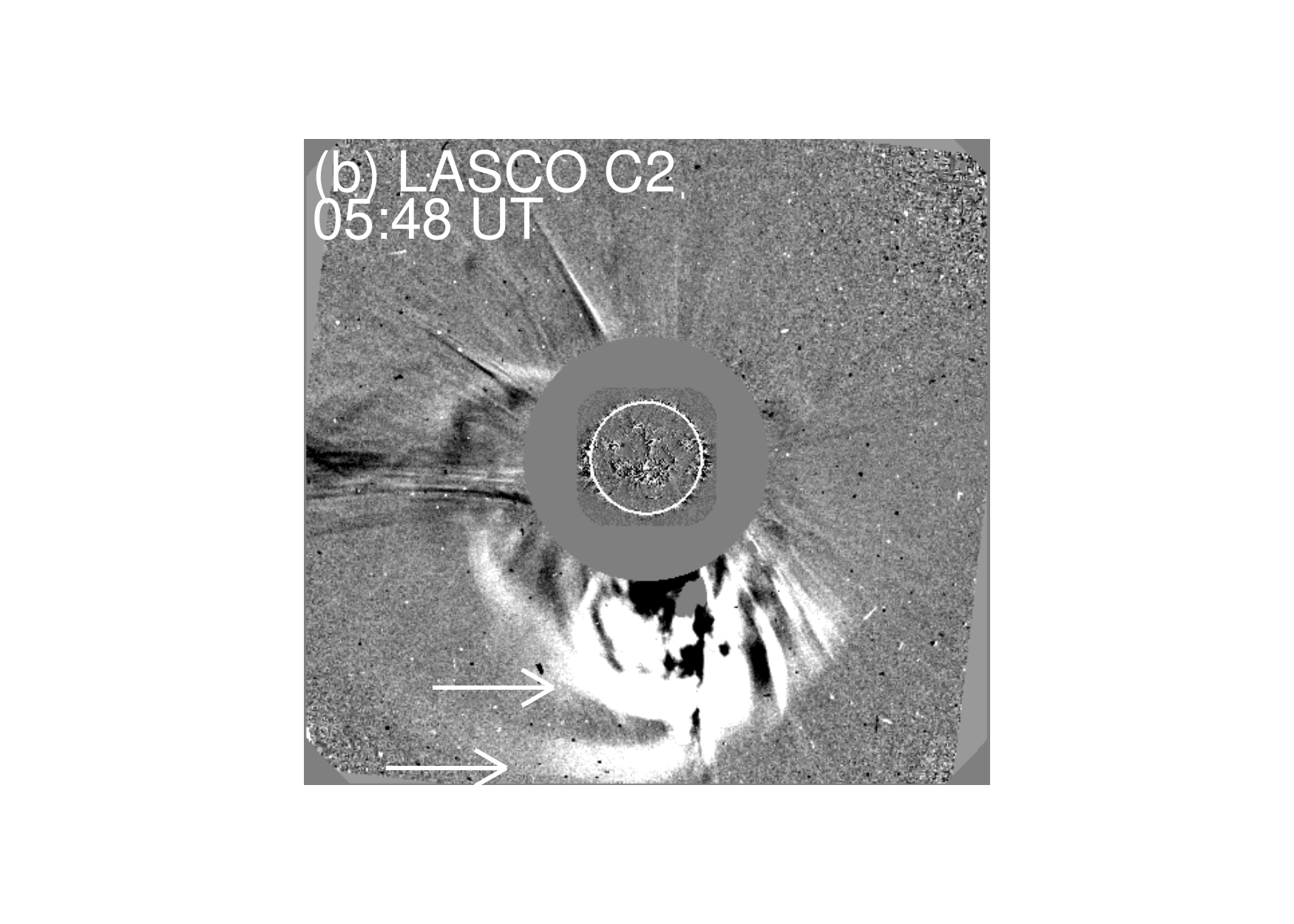}
	}
\vspace*{-3.3cm}
\centerline{
	\hspace*{0\textwidth}
	\includegraphics[width=0.852\textwidth,clip=]{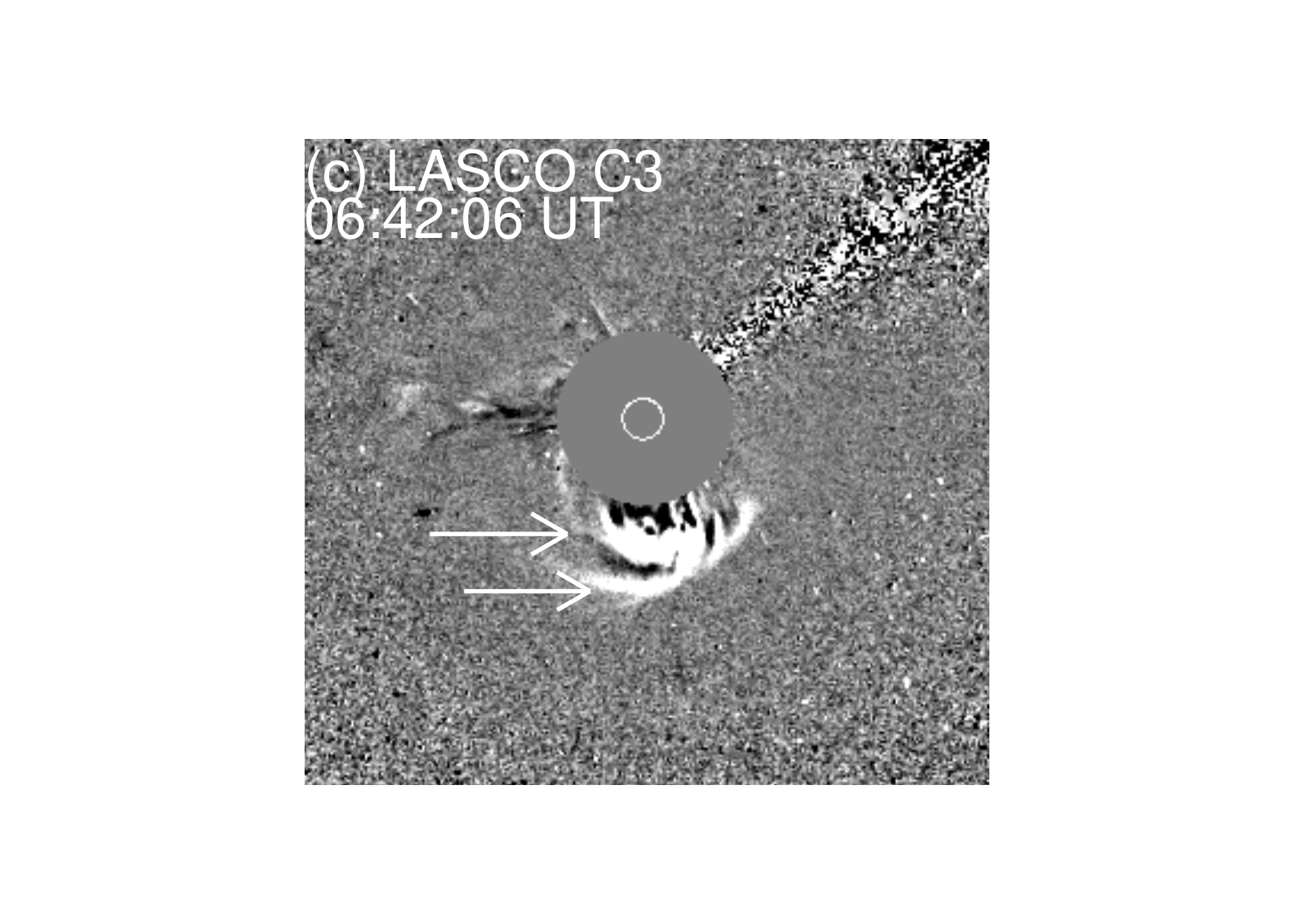}
	}
\vspace*{-1.5cm}
\caption{LASCO C2 ((a)--(b)) and C3 (c) coronograph white light images of the CME associated with the eruptive flare event. The difference images have been downloaded from \url{https://cdaw.gsfc.nasa.gov/CME\_list/}. The arrows represent the two parts of the leading edge of the CME.}
\label{fig11}
\end{figure}

The dynamic evolution of the circular and parallel ribbons can be seen in the AIA 304 and 1600 \AA\ images (see Figures~\ref{fig8}(b)--(f) and~\ref{fig9}(b)--(f), respectively). It is noteworthy that by $\approx$04:33 UT the strong emission from parallel ribbons overtakes the emission coming from larger circular ribbon structure (Figures~\ref{fig8}(c), and~\ref{fig9}(d)). These parallel ribbons appear to grow in size and also to separate away from each other (see Figures~\ref{fig8}(e)--(f) and~\ref{fig9}(e)--(f) and associated animations). Simultaneous to the growth of parallel ribbons, the circular flare ribbons continue to exhibit enhanced emission. Also, the RB remains prominent during this phase while it slowly extends westward with time as well (see Figures~\ref{fig8}(c)--(f), Figure~\ref{fig9}(d)--(f) and associated animation). This observational feature can also be infered from the intensity plot. At $\approx$04:28 UT, we see another stage of intensity inhancement (shown by the vertical blue bashed line in Figures~\ref{fig6}(a)--(c)) in all the the circular, parallel and, remote brightening regions. This is due to the fast eruption of flux rope, which intensify the reconnection (both null-point and standard) and hence the intensity of circlar, parallel and, remote ribbons.

The eruptions from the AR eventually produce a halo CME which is observed by LASCO on--board the {\it SOHO} (Figure~\ref{fig11}). The CME is first observed by LASCO C2 at a height of 3.25 $\rm R_\odot$ at $\approx$05:00 UT. We note that the CME exhibits a complex structure with the clear appearance of two distinct yet connected leading edges (shown by the arrows in Figure~\ref{fig11}). According to the LASCO CME catalog \footnote{$https://cdaw.gsfc.nasa.gov/CME\_list/UNIVERSAL/2014\_12/htpng/20141217.050005.p162g.htp.html$} the CME propagates within LASCO C2 and C3 field--of--view with a linear speed of $\rm \approx$587 $\rm km~s^{-1}$.

\begin{figure*}
\vspace*{1cm}
\centerline{
	\hspace*{0\textwidth}
	\includegraphics[width=1\textwidth,clip=]{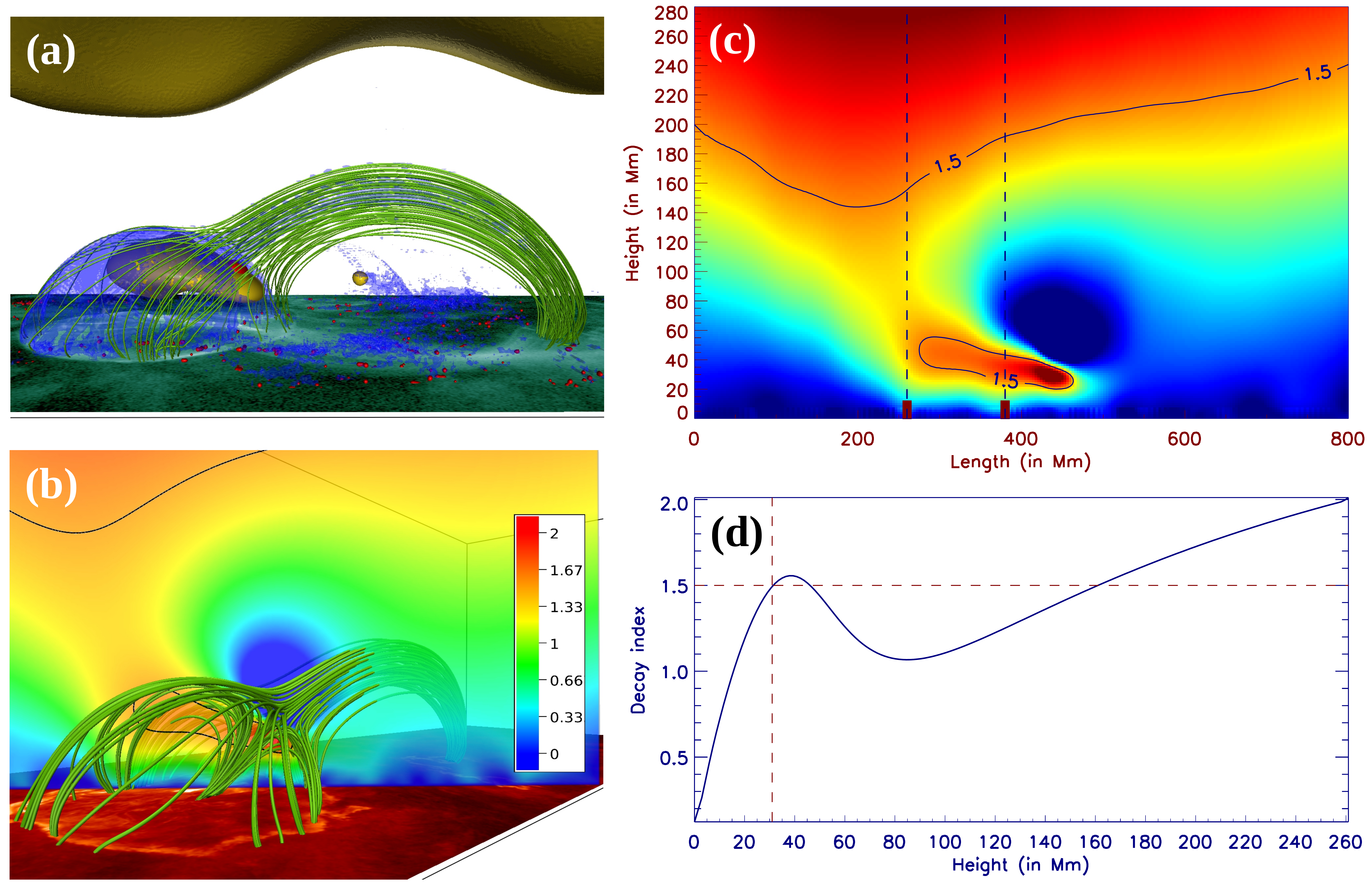}
	}
\vspace*{0cm}
\caption{(a)--(d) Distribution of decay index over the active region NOAA 12242. In panel (a) we plot an isosurface of n=1.5 by yellow color. For comparison, we have also indicated the region of the null-point in brown color. Two regions (inside and outside the fan dome) have been observed with decay index value equal or greater than 1.5. The blue coloured iso-surface shows the QSL (log(Q)=5), which represents the fan surface. (b) Distribution of decay index along the vertical surface that passes through the parallel ribbons. The black curves on the vertical surface indicate n=1.5.  For further clarification, we have separately plotted the surface in the panel (c). We have approximated the locations of the flare ribbons by the two solid brown boxes near the bottom boundary of the plot (see panel (c)). The average values of decay index (along this surface) as a function of height within the two vertical dotted lines are plotted in the bottom panel (d). From this panel, we find that the decay index reached the value of 1.5 at a height of $\approx$30 Mm.}
\label{fig12}
\end{figure*}

In order to determine whether the torus instability is the triggering mechanism of the flux rope eruption, we have calculated the distribution of magnetic decay index over the AR volume (i.e., $\rm 510\times 180\times 180$ pixels) using potential magnetic field extrapolation (see Figure~\ref{fig12}). Decay index (n) is calculated using the relation: $\rm n=-d[log(B_{ex})/d[log(h)]$; where $\rm B_{ex}$ and h are potential external field and height, respectively \citep{Bateman78,kliem06}. Figure~\ref{fig12}(a) represents closed 3D null--point magnetic configuration over the AIA 304 \AA\ image with iso--surfaces of n=1.5 shown by yellow color. As the image suggests, there are two regions in the AR volume where the value of decay index is greater than 1.5: inside the fan-dome (within the closed iso--surface) and in the high coronal altitude (the open iso--surface at the top of the figure). While the open iso--surface at the top is rather standard, the closed surface inside the dome is important and useful in our case. It suggests us that the majority of the upper region within the fan-dome was subjected to torus instability. In Figure~\ref{fig12}(b), we have drawn a vertical
\newpage
\noindent surface that passes through the parallel ribbons, where the distribution of decay index along the surface is plotted. The black curves on the vertical surface indicate n=1.5. For further clarification, we have separately plotted the surface in  Figure~\ref{fig12}(c). We have approximated the locations of the flare ribbons by the two solid brown boxes near the bottom boundary of the plot. The average values of decay index (along this surface) as a function of height within the two vertical dotted lines are shown in Figure~\ref{fig12}(d). From this panel, we find that the decay index reached the value of 1.5 at a height of $\approx$30 Mm. 

We would like to emphasize that the n=1.5 region inside the dome directly borders the fan surface. To explore this further, we calculated the QSL in the flaring region which also includes the fan-dome and depicted the high Q-regions (log (Q)=5) by blue iso-surface (Figure~\ref{fig12}(a)). Our analysis readily suggest that the flux rope activation by the torus instability would immediately lead to reconnection at the null point.


\subsection{{\bf \it RHESSI} X--ray observations}
\label{sec4.3}

The {\it RHESSI} observations are available during some part of the peak and decay phases of the M--class flare while it completely missed its rise phase (Figure~\ref{fig4}(a)). We also find useful {\it RHESSI} observations during a part of the pre--eruptive phase. Some of the {\it RHESSI} imaging observations during the pre--eruptive phase are shown in Figures~\ref{fig5}(b)--(c) where the background corresponds to the AIA 131 \AA\ images. It is noteworthy that {\it RHESSI's} attenuator state was A0 during this phase, i.e., {\it RHESSI} observed with its highest sensitivity at low energies. This has enabled us to examine the X--ray sources associated with the weak pre--flare emission down to 3 keV energy. We note that the earliest X--ray emission at 3--6 keV and 6--12 keV energy bands originates from the location of pre--flare brightening (Figure~\ref{fig5}(b)). At this time, the main source presents a broad structure while secondary sources, situated south of the main source, are also noticeable. In the later stages of the pre--eruptive phase, the X--ray sources converge toward the bright loop--like structure which is clearly noticeable in the background EUV images (Figure~\ref{fig5}(b)--(c)). The {\it RHESSI} observations reveal that the pre--flare X--ray emission is persistent and spatially not so--localized. From the comparison of X--ray and EUV images along with the corresponding magnetogram, we infer that the pre--flare brightening is suggestive of the slow evolution of a bundle of field lines that continues to emit enhanced hot emission while undergoing upward expansion. Based on temporal and spatial characteristics of pre--flare emission over a wide range of energy scales, we argue that it is related to the slow magnetic reconnection, presumably of tether--cutting category.

\begin{figure*}
\vspace*{-3.8cm}
\centerline{
	\hspace*{0.05\textwidth}
	\includegraphics[width=0.61\textwidth,clip=]{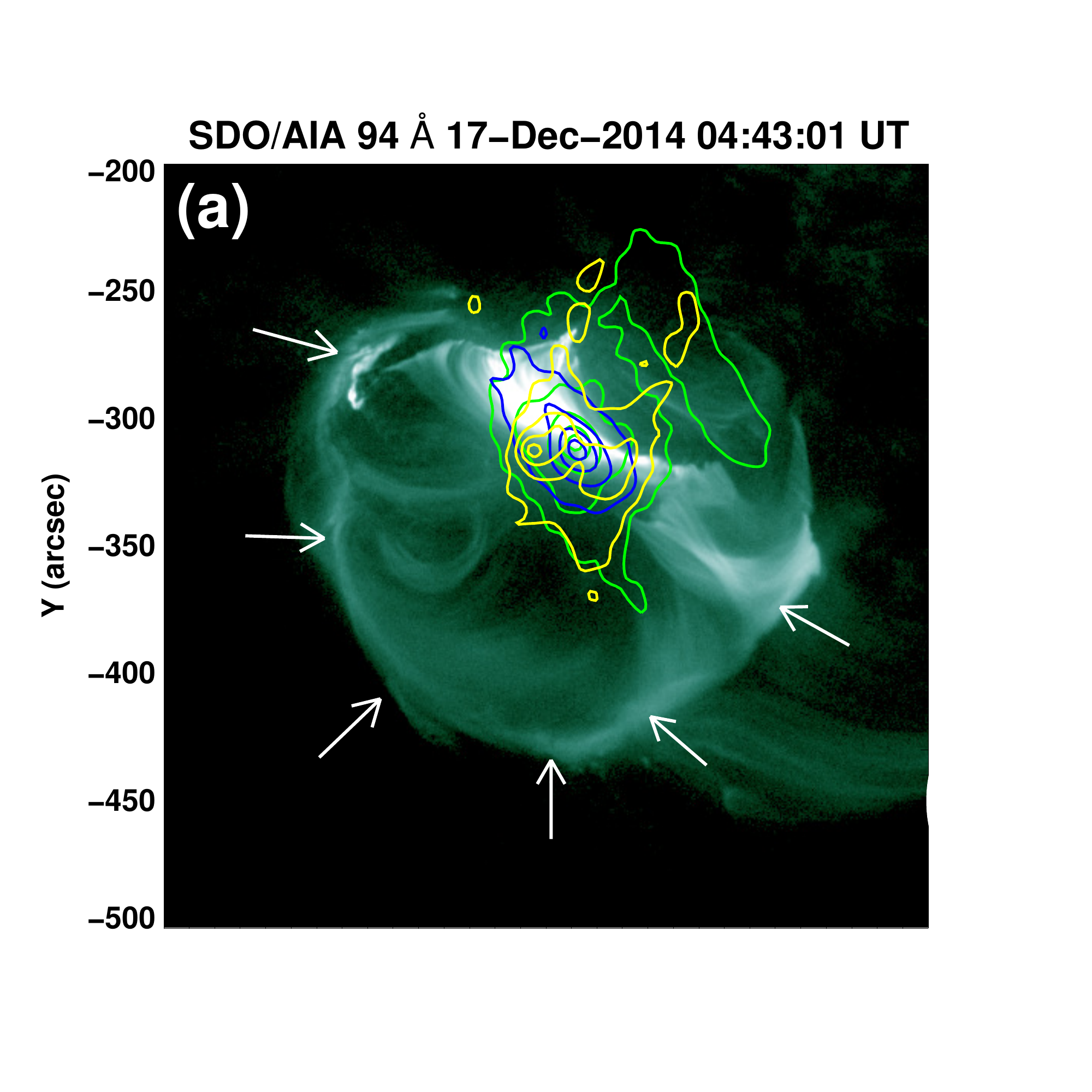}
	\hspace*{-0.171\textwidth}
	\includegraphics[width=0.61\textwidth,clip=]{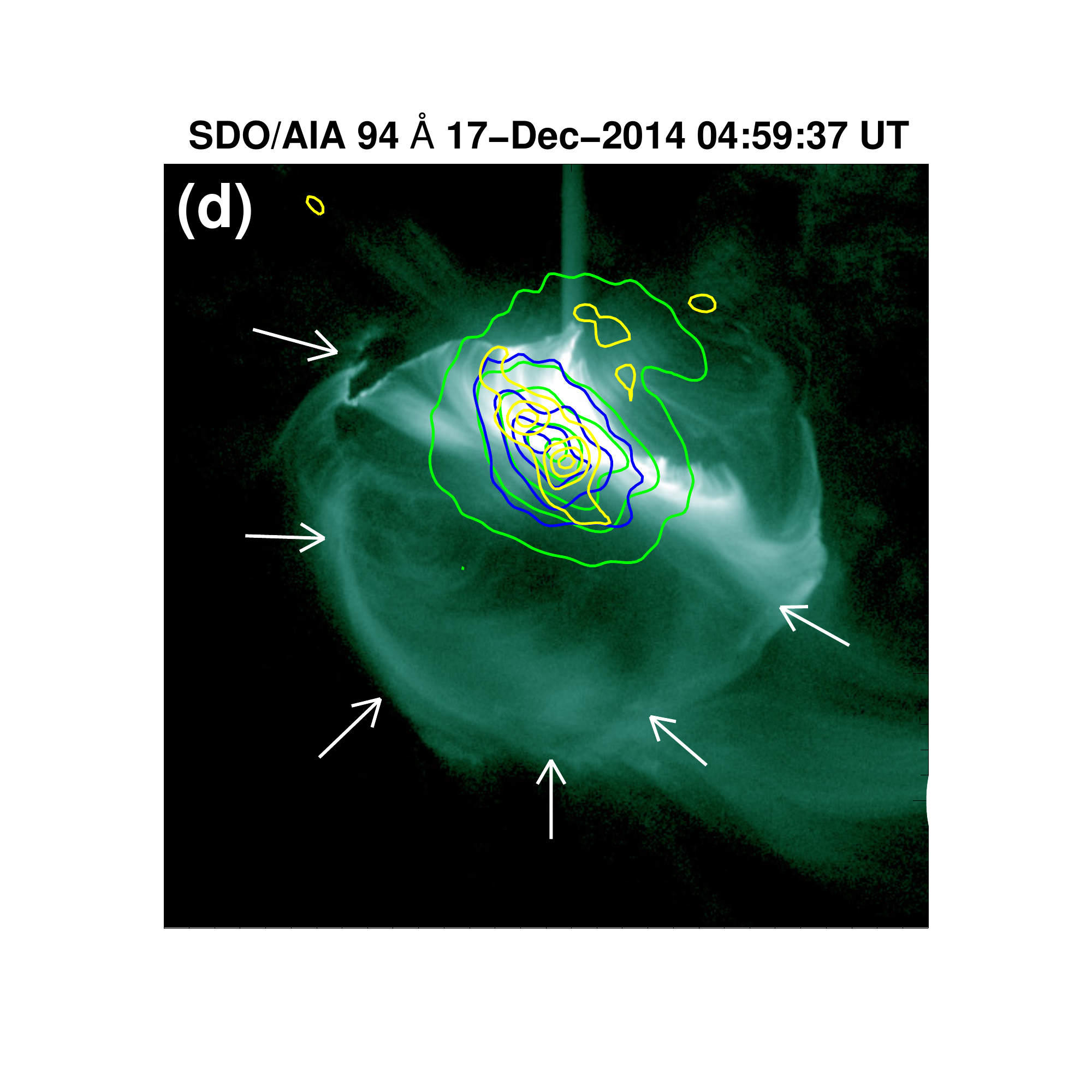}
	}
\vspace*{-2.6cm}
\centerline{
	\hspace*{0.05\textwidth}
	\includegraphics[width=0.61\textwidth,clip=]{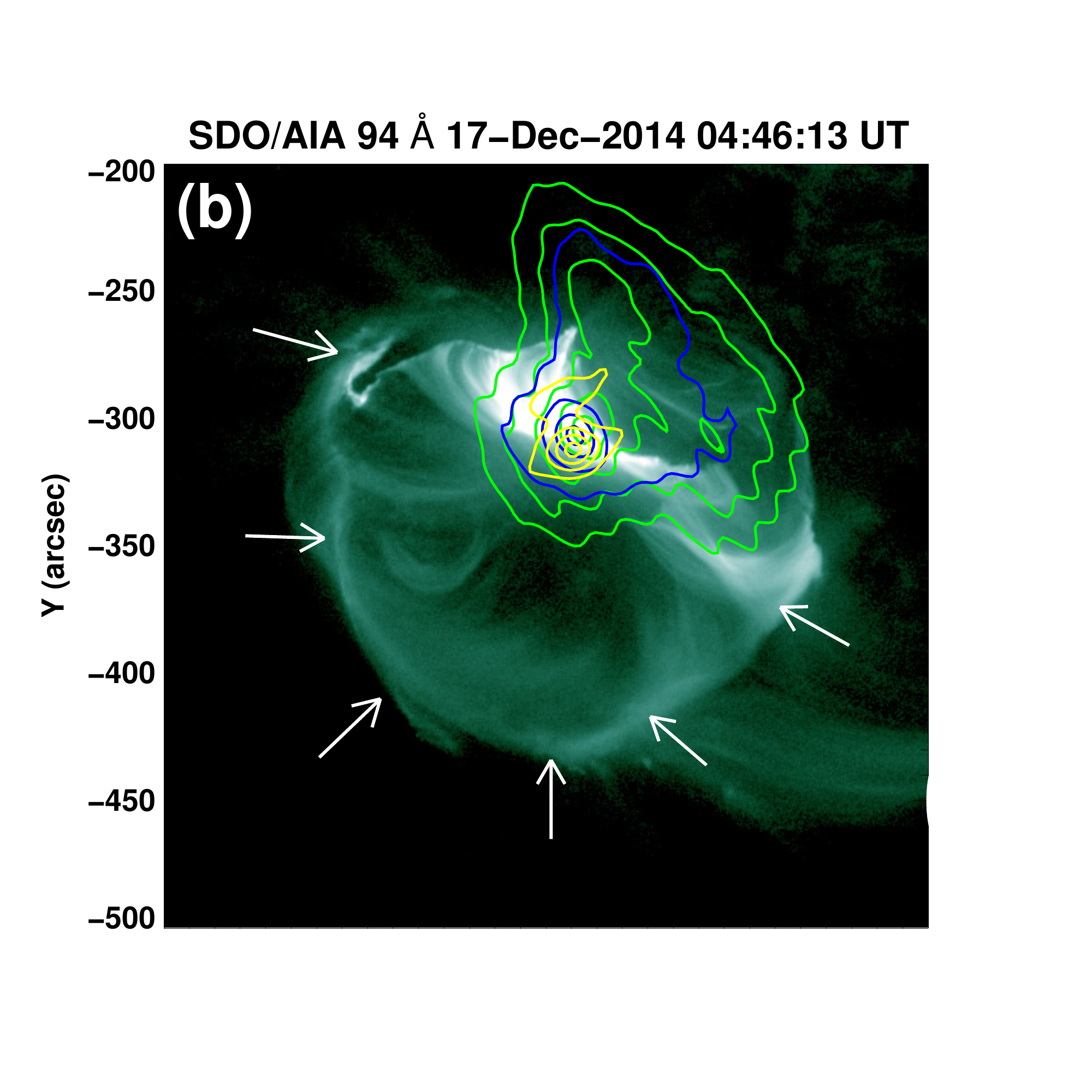}
	\hspace*{-0.171\textwidth}
	\includegraphics[width=0.61\textwidth,clip=]{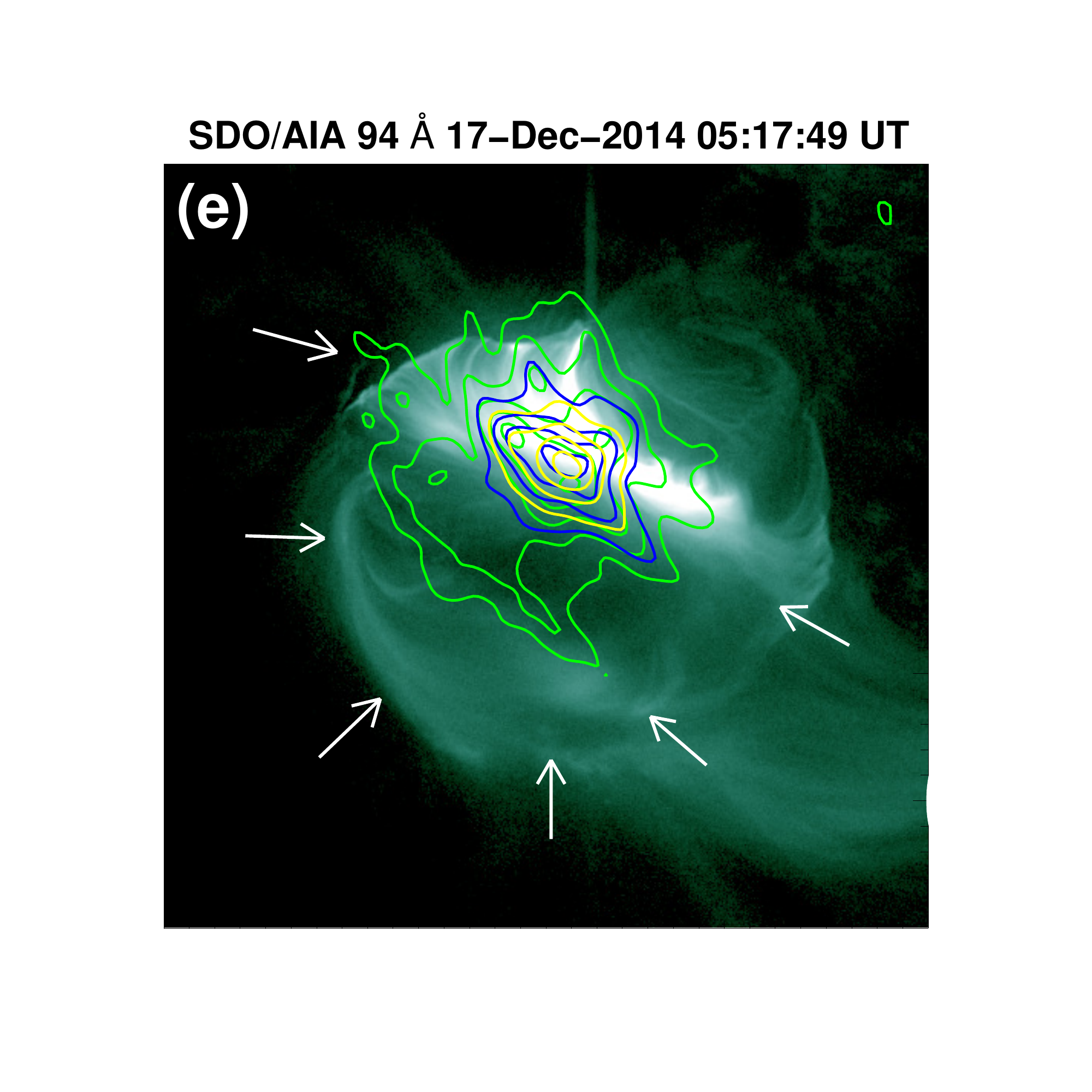}
	}
\vspace*{-2.6cm}
\centerline{
	\hspace*{0.05\textwidth}
	\includegraphics[width=0.61\textwidth,clip=]{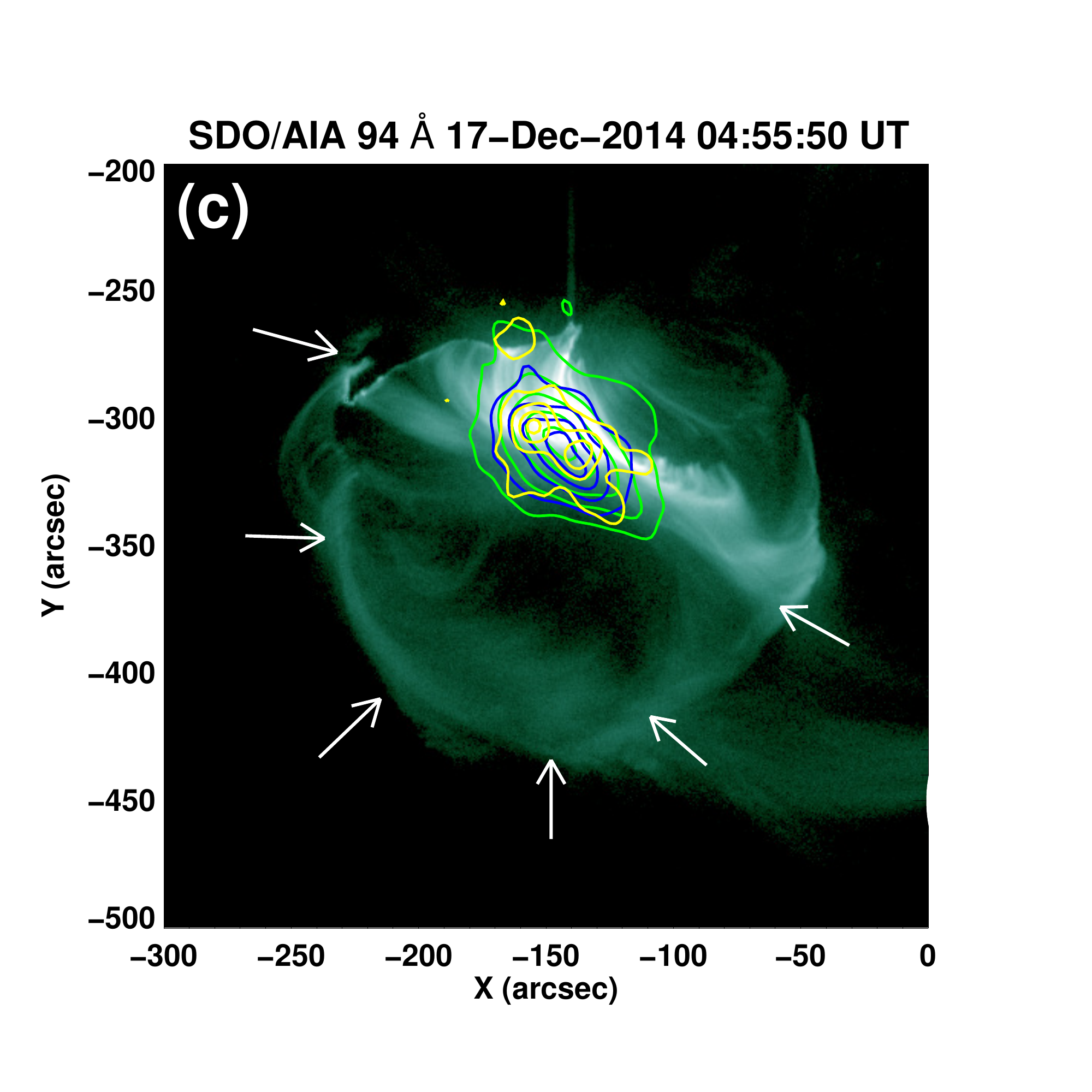}
	\hspace*{-0.171\textwidth}
	\includegraphics[width=0.61\textwidth,clip=]{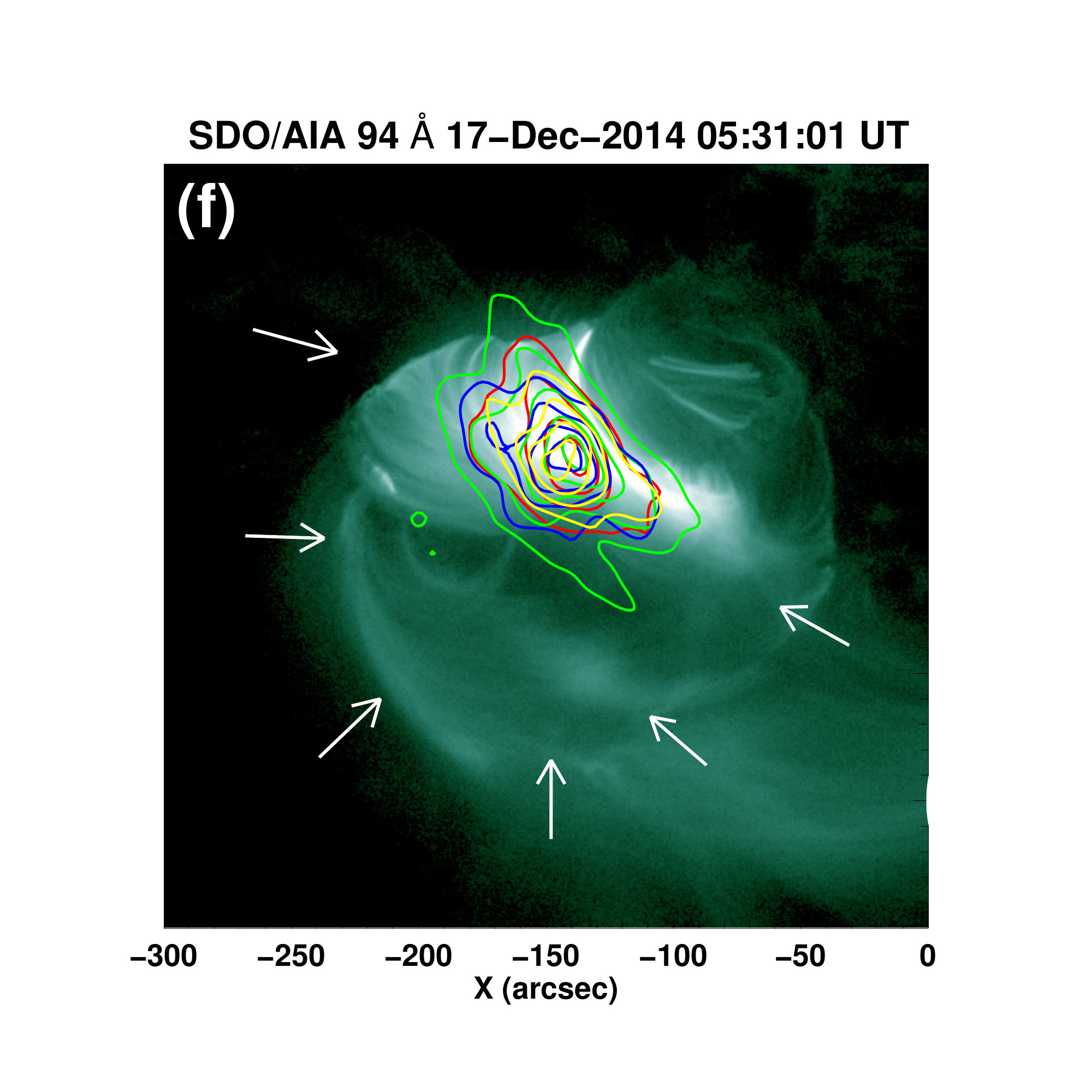}
	}
\vspace*{-0.9cm}
\caption{Representative {\it RHESSI} X--ray images overplotted on AIA 94 \AA\ images. The red, green, blue and yellow contours represent the RHESSI X--ray contours at 5--10, 10--15, 15--25, and 25--50 keV energy bands, respectively. Contour levels are 35\%, 60\%, 80\%, and 95\% of the peak intensity for 5--10 keV, 15--25 keV, and 25--50 keV sources. To examine the extent of hot plasma emission, we set the contour levels as 20\%, 40\%, 60\%, 80\%, and 95\% of the peak intensity for 10--15 keV sources. Arrows represent the edge of the circular structure.}
\label{fig13}
\end{figure*}

In Figure~\ref{fig13} we show the evolution of X--ray sources with respect to EUV 94 \AA\ flaring image in four representative energy bands, namely, 5-10, 10-15, 15-25, and 25-50 keV. The X--ray source in $\lesssim$15 keV are mostly dominated by the thermal emission from the hot flaring loops while at higher energies ($\gtrsim$20 keV) the non--thermal component from foot--point regions becomes predominant.
Notably, during this whole phase the centroids of X--ray sources at lower and higher energies are almost co--spatial, thus indicating the presence of multi--thermal plasma within a confined environment. Further, at times, the X--ray emitting sources exhibit very extended structures (see panels (a), (b), (d) and (e) of Figure~\ref{fig13}), mainly at 10--15 keV energy band. We also emphasize that, in contrast to the standard flare scenario, the X--ray sources here do not undergo much spatial evolution. Comparison of X--ray and EUV images reveals that the broad X--ray sources likely indicate the composite thermal emission caused by the reconnection of magnetic field lines spread in a larger yet confined region i.e., the region is delineated by a 3D null--point magnetic topology. In Figures~\ref{fig8}(e)--(f), we compare the spatial distribution of EUV and X--ray sources in the light of the formation of parallel ribbons. From the co--spatial images, it is evident that the compact non-thermal source (i.e. 25-50 keV emission) originated mainly from the location of southern flare ribbon and the region in-between the parallel ribbons. This suggests asymmetry in the evolution of HXR footpoint sources. Thus, the spatial structures associated with 25-50 keV sources suggest that the non-thermal emission observed during this circular flare came from the footpoints of short loops lying within the inner PIL (Figure~\ref{fig2}(a)) by the standard flare reconnection. On a whole, the structure and evolution of X--ray sources present a complex situation associated with the release of the magnetic energy by the reconnection in 3D null as well as standard flare reconnection in the underlying magnetic configurations. 

\begin{figure*}
\vspace*{-1cm}
\centerline{
	\hspace*{0\textwidth}
	\includegraphics[width=1\textwidth,clip=]{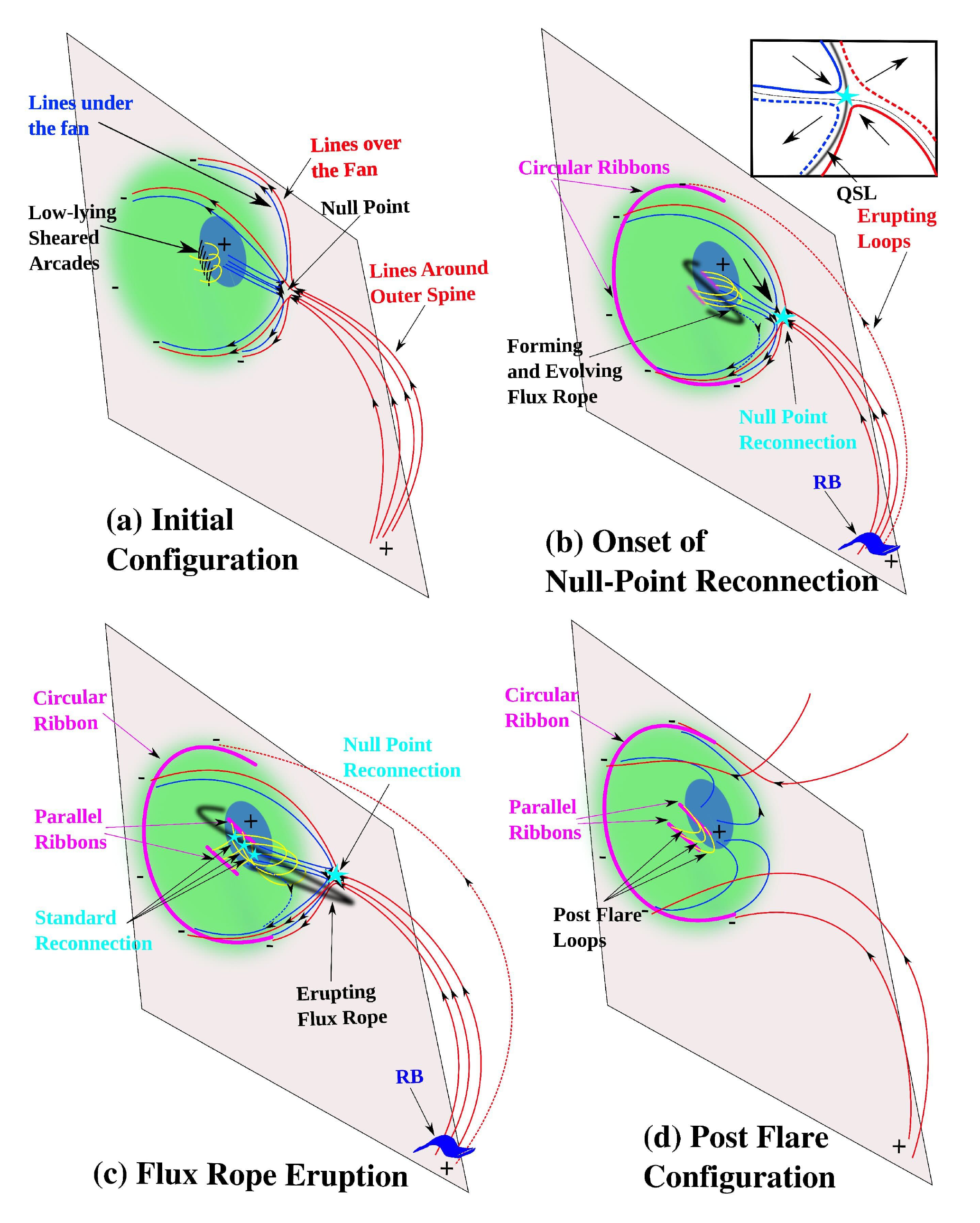}
	}
\vspace*{0.1cm}
\caption{Schematic representation of the different phases of the eruptive flare event. The field lines enclosed below and over the fan are represented by the blue and red colors, respectively. The remote brightening (RB) is represented by the blue color structure in panels (b) and (c). Reconnection regions are marked by the cyan color stars in panels (b) and (c). The circular and parallel flare ribbons are shown by the solid purple lines in panels (b)--(c). Flux rope is shown by the black line in panels (b)--(c).}
\label{fig14}
\end{figure*}


\section{Results and Discussion}
\label{sec5}

\begin{table*}
 \caption{Observational Sequence of the Event}
 \label{Table1}
 \begin{tabular}{ll}
  \hline
Time & Observational features \\
\hline
$\approx$03:45 UT & Appearance of circular ribbon in 304 \AA\ images along with remote brightening.\\
 & -Starting of compact brightening at the inner PIL region.\\
$\approx$04:18 UT & Start of flare's impulsive phase; formation of parallel ribbons started; \\
 & Circular ribbon continued to appear with impulsive increase in intensity.\\
 & -Onset of the eruption of overlying loops observed in hot 94 \AA\ channel.\\
$\approx$04:25 UT & Onset of the inner field lines (either flux rope of sheared arcades) observed in hot 94 \AA\ channel. \\
$\approx$04:51 UT & Peak phase of the flare observed from GOES X-ray flux profile. \\
$\approx$05:00 UT & First appearance of CME observed in LASCO-C2 field of view. \\
\hline
\end{tabular}
\end{table*}

In Figure~\ref{fig14}, we provide schematic diagrams to explain and interpret the multi--phase evolution of the complex circular ribbon flare exhibiting distinct parallel conjugate ribbons within the periphery of the circular ribbon. For the convenience, we summarize the chronology of the various evolutionary phases of the event in Table~\ref{Table1}. The magnetic topology over the AR is of closed 3D null--point topology where a bipolar region lies inside the fan dome (cf. Figures~\ref{fig14}(a) and~\ref{fig2}). This type of magnetic configuration consists of field lines enclosed below the fan and over the fan (shown by blue and red lines, respectively in Figure~\ref{fig14}(a)) with a magnetic null and a distinct region showing RB \citep[e.g.,][]{Sun13,Joshi17}. A flux rope evolving inside the fan--dome (Figure~\ref{fig5}) is likely formed due to the tether--cutting reconnection caused by the magnetic flux changes at the inner PIL (see Figure~\ref{fig3}). In this context, we would like to refer some studies which suggest that the mechanism responsible for such pre--flare brightenings for the formation of flux ropes and flare arcades are the result of magnetic reconnection at the hyperbolic flux tube (HFT) which is the topological element below a forming flux rope by assuming that the observed structure is a flux rope \citep[see e.g.,][]{Aulanier10,Janvier13}. We propose that the slowly evolving flux rope pushed the lines enclosed below the fan lines toward the magnetic null and trigger the null--point reconnection. In this scenario, the reconnection region would lie between the inner blue fan lines and the outer red fan lines which is depicted in Figure~\ref{fig14}(b), where the location of the null--point reconnection is shown by sky blue colored star in Figures~\ref{fig14}(b)--(c). As a consequence of null point reconnection, erupting loops are observed (cf. Figures~\ref{fig14}(b) and~\ref{fig7}(a)--(c)) which likely represent newly formed loop system moving away from the null-point. Thus, the null point reconnection would facilitate the eruption of the underlying flux rope lying within the fan dome. The next phase starts with the onset of flux rope eruption and standard reconnection occurs underneath it (cf. Figures~\ref{fig14}(c) and~\ref{fig7}(d)--(e)). This reconnection enhances the brightness of two parallel ribbons that form on both sides of the inner PIL inside the fan dome (cf. Figures~\ref{fig14}(c),~\ref{fig8}(a)--(f), and~\ref{fig9}(a)--(f)). In the 3D model for the flux rope eruption, the standard reconnection corresponds to magnetic reconnection in the HFT behind the erupting flux rope \citep[see][]{Aulanier10,Janvier13}. The null--point reconnection continues with the eruption of the underlying flux rope until null--point topology gets partially disrupted. The configuration just after the eruption is shown in Figure~\ref{fig14}(d). Later on, the system gets back to a potential stage with 3D null--point magnetic configuration, which is similar to the initial configuration (Figure~\ref{fig14}(a)). 

Recently, \cite{Joshi15} and \cite{Joshi17} studied flares that show two different sets of flare ribbons viz. parallel and circular in an AR with a 3D null--point magnetic topology. They observed that the parallel ribbons form before the circular ribbon. They suggested two stages of magnetic reconnection to explain their observations like: initial standard followed by later null--point reconnection. In this paper, we present a reverse case in which the signatures of circular ribbon comes first in the pre--flare phase. The parallel ribbon starts to form in the implusive phase. Both kinds of ribbons underwent rapid intensity enhancements as the flux rope activated towards eruption.

Flares that consist of a circular ribbon are usually found to occur in a typical 3D null--point topology via the null--point reconnection \citep{Aulanier00,Jiang14,Sun13,Joshi15,Joshi17}. In this work, we provide further insights of the different reconnection stages possible in the null--point topology by carefully analyzing the temporal aspects in the developments of circular and parallel ribbon flares. We observed a slow rise in the brightness of the circular ribbon area at $\approx$03:45 UT in AIA 304 \AA\ images (see Figure~\ref{fig6}(a)). The apparent eruption of overlying loops started at $\approx$04:18 UT (Figures~\ref{fig7}(a)--(c),~\ref{fig10}), which is co--temporal with the onset of sharp enhancement in the circular ribbon brightening and RB (see Figures~\ref{fig8}(a)--(b) and animations associated with Figures~\ref{fig8} and~\ref{fig9}). Circular morphology of the flare ribbon observed in the early phase and magnetic configuration inferred from NLFFF extrapolation signify magnetic reconnection in the null point topology. As a consequence of this reconnection, two distinct flux systems below and above the fan separatrix would reconnect (see schematic shown in Figure~\ref{fig14}(b)). This would create a newly reconnected overlying hot coronal erupting loops (Figures~\ref{fig7},~\ref{fig10} and~\ref{fig14}). According to our understanding the erupting (overlying) loops observed from $\sim$04:10 UT to $\sim$04:25 UT (Figure~\ref{fig7}(a)-(c) and corresponding measurements in Figure~\ref{fig10}) presumably trace the newly formed hot loops moving away from the null point. The eruptive expansion of overlying loops caused due to the null-point reconnection imply removal of the constraining flux that envelop the flux rope. This proceeds toward decreasing the downward tension of the restraining fields and hence ease the eruption process of the underlying flux rope. Afterwards, the field lines continue to reconnect across the QSL--halo surrounding the outer spine. The illustration showing the distribution of QSL along with torus unstable region within the fan-dome (Figure~\ref{fig12}(a)) further clarifies the eruption scenario. The fact that the torus unstable region (i.e., region with n=1.5) inside the dome directly borders the fan surface (having large Q-values) confirms that any instability of the flux rope would immediately trigger reconnection at the fan-QSL leading the circular ribbon brightening. After the reconnection, the field lines may appear to slip within the same dome and hence the sequential circular brightening \citep[][]{Sun13,Joshi15,Joshi17}. In our case, we observed counterclockwise sequential movement of brightening in a circular manner (see animation associated with Figure~\ref{fig9}), which provide evidence of the reconnection at QSL.

We would like to elaborate that during the evolution of the present event, the role of null--point reconnection toward the eruption of the underlying flux rope in null--point topology (evident from the initial circular ribbon brightening) seems to match with the null--point reconnection in the breakout model of solar eruption \cite[]{Aulanier00,Sun13,Liu15,Xu17,Li18}. However, we also state that the null--point reconnection alone may not be able to explain the eruption of underlying quasi--stable flux rope. The NLFFF analysis suggests the location of null--point to be at $\rm \approx 36~Mm$ which implies that the  null--point reconnection occurred at lower coronal heights. This interpretation is also supported by the fact that (1) the mild yet persistent circular brightening during the pre--flare phase (see Figure~\ref{fig6} and animation associated with Figures~\ref{fig8} and~\ref{fig9}); (2) cotemporality between the eruption of overlying loops and sharp enhancement of circular as well as RB (cf. Figures~\ref{fig8}(a)--(b),~\ref{fig9}(a)--(b) and Figure~\ref{fig10}); (3) the time delay ($\approx$7 min) between the onset of eruption of the underlying flux rope ($\approx$04:24 UT; Figure~\ref{fig10}) and the overlying filed lines ($\approx$04:17 UT; Figure~\ref{fig10}; see Figure~\ref{fig7} and associated animation). This fact is supported by the observations that a sharp increase in the brightness of the circular ribbon (a proxy for the increase in the rate of null-point reconnection) occurred simultaneously with the observed eruption of overlying loops (cf. Figures~\ref{fig6}(a),~\ref{fig7} and~\ref{fig10}). We suggest that the transfer of flux across the fan separatrix (Figure~\ref{fig14}(b)) observationally inferred from erupting loops would contribute in alleviating the immediate constraint on the core field. This would facilitate the subsequent eruption of underlying flux rope (cf. Figures~\ref{fig7}(a),~\ref{fig8}(a) and~\ref{fig10}). The removal of overlying flux as a cause of the eruption of the underlying flux rope at different magnetic configurations has also been investigated in earlier studies \citep[e.g.,][]{Joshi13b,Cheng13,Sterling14}. The decay index analysis suggests that the torus instability may also play an important role in triggering the eruption of the underlying flux rope. The instability zone (i.e., region of decay index n$\gtrsim$1.5) lies at a height of $\approx$30 Mm (see Figure~\ref{fig12}). When the flux rope reached at this height within fan dome, it attained eruptive motion \citep[see similar results by][]{Jiang13,Jiang14}. We also note that, once the flux rope reached the region above the null point, the value of corresponding decay index decreased below 1.5 which points toward the stability zone within a height range in the higher coronal level. However, the CME observations clearly reveal the successful eruption of the activated flux rope without any further hindrance. We, therefore, see that the triggering processes of the flux rope eruption with eruptive flare, initiated below a null could be very complex with several factors being involved. Our comprehensive multi-wavelength data analysis and coronal magnetic field modeling suggest that, both the torus instability and breakout model can contribute to the triggering mechanism for this eruptive flare. Hence, for eruptive flares initiated below a null--point, more observational and theoretical studies need to be done to determine the exact triggering mechanism.


\section{Conclusions}
\label{sec6}

In this work, we present a multi--wavelength analysis of an interesting multiple ribbon solar flare that occurred on 2014 December 17. The flare occurred in a typical 3D null--point topology. The multi--wavelength observations clearly reveal three distinct phases in the evolution of this solar eruptive flare. The main results of this study are summarized below.

\begin{enumerate}
\item Prior to the onset of eruption (i.e., pre--eruptive phase), we find evidence for the formation and slow expansion of the flux rope which is corroborated by the enhanced EUV emission from an expanding loops system and spatially extended low energy X--ray sources $\rm (\lesssim 12~keV)$. The subtle yet obvious manifestation of the circular ribbon along with the anemone shape of the fan field lines during the pre--eruptive phase imply the onset and persistence of null--point reconnection well before the later standard flare reconnection. Simultaneous to the commencements of circular ribbon, we also observed RB which overall agree with the close null--point configuration with the outer spine as inferred from the extrapolations. 
\item In the impulsive phase, the flare underwent the dynamic restructuring of the field lines below and above the fan via null--point reconnection along with the eruptive expansion of the outer (overlying) hot coronal loops. The eruption of the overlying fields would remove the constraining flux, thereby facilitating the eruption of the underlying flux rope. The decay index analysis suggests that the flux rope resided close to an unstable zone which starts at a height of $\approx$30 Mm inside the fan dome. When the evolving flux rope would reach this height, it would undergo eruptive expansion. Here we note that the decay index decreases above the null point which may inhibit the further expansion of flux rope. However, the combination of \textit{SDO}/AIA and \textit{SOHO}/LASCO observations clearly reveal the successful eruption of the flux rope without any interruption in the overlying coronal layers. 
In our case, we propose that the combination of coronal breakout and torus instability within the fan-spine dome facilitated  the successful eruption of the flux rope resulting the large M--class flare and a CME.
\item In the subsequent stages, the characteristics of the eruptive flare match well with the classical picture of the two--ribbon flare. Now, flare evolved with the complete expulsion of the underlying flux rope, which subsequently gave rise to a CME.
\end{enumerate}

This event is a nice example of the destabilization and eruption of a flux rope initially confined in a null--point topology. More such observational studies and corresponding numerical simulations are needed to clearly understand the eruption scenario in complex circular--cum--parallel ribbon flares.



\section*{Acknowledgements}

We thank \textit{SDO}, \textit{RHESSI}, \textit{SOHO} and \textit{GOES} teams for their open data policy. \textit{SDO} is NASA's missions under Living With a Star (LWS) and \textit{RHESSI} was NASA's missions under SMall EXplorer (SMEX) programs. \textit{SOHO} is a project of international cooperative effort between ESA and NASA. We also thank Dr. Thomas Wiegelmann for providing the NLFFF code used in the analysis. We thank Dr. Jaroslav Dud\'{i}k, the reviewer of our paper, for insightful comments that improved the scientific content of the paper.

\section*{Data Availability}
Observational data from AIA and HMI on board \textit{SDO} utilised in this article are available at \url{http://jsoc.stanford.edu/ajax/lookdata.html}. X-ray observations from \textit{RHESSI} are available at \url{http://hesperia.gsfc.nasa.gov/hessidata/}. The NLFFF code employed in this article for coronal magnetic field modelling is provided by Dr. Thomas Wiegelmann. Different aspects of the code are explained and discussed in \url{https://doi.org/10.1023/B:SOLA.0000021799.39465.36}, \url{https://doi.org/10.1007/s11207-006-2092-z}, \url{https://doi.org/10.1051/0004-6361/201014391}, \url{https://doi.org/10.1007/s11207-012-9966-z}. The IDL-based code used for the computation of $Q$ and $T_w$ is available at \url{http://staff.ustc.edu.cn/~rliu/qfactor.html}.

\section*{Supplementary Material}
Videos are attached with Figures \ref{fig3} (Movie-Figure3.mp4), \ref{fig5} (Movie-Figure5a.mpeg, Movie-Figure5b.mpeg), \ref{fig7} (Movie-Figure7.mpeg), \ref{fig8} (Movie-Figure8.mpeg) and \ref{fig9} (Movie-Figure9.mpeg) which are available in the online article.



\bibliographystyle{mnras}
 \newcommand{\noop}[1]{}




\end{document}